# Experimental Study on the Detection of Frozen Diffused Ammonia Blockage in the Inactive Section of a Variable Conductance Heat Pipe

F. K. Miranda [a,b], Romain Rioboo [a], Mikael Mohaupt [a], Cristian Marchioli [b,c,1]

[a] *Euro Heat Pipes SA, 1400 Nivelles, Belgium*
[b] *University of Udine, Department of Engineering and Architecture, 33100 Udine, Italy*
[c] *International Centre for Mechanical Sciences (CISM), Department of Fluid Mechanics, 33100 Udine, Italy*

## *Abstract*

Variable Conductance Heat Pipes (VCHP) are mainly employed to cool down electronic systems in spacecraft applications, as they can handle high temperature fluctuations in their cold source, preventing thus the systems from damaging. These fluctuations, as well as ultra-low temperatures, are always present in outer space, and one of the key steps in a VCHP design is therefore to make sure that they endure these conditions. However, not much has been written about their resilience during and after a long exposition to subfreezing conditions, i.e. temperatures lower than the freezing point of the working fluid. In this paper we implement and validate a computational routine based on a modified Flat-Front Approach to predict the VCHP temperature profile and to determine the location of the gas-vapor front. Then we continuously expose an ammonia/stainless-steel VCHP to temperatures below the ammonia freezing point for 211 hours, to later examine the formation and subsequent dynamics of a thin block of frozen ammonia which is diffused into the inactive part of the heat pipe condenser. We describe as well how a strong correlation between the adiabatic section and the reservoir temperatures is maintained (or broken) upon the occurrence (or absence) of the blockage of frozen ammonia.

*Keywords — VCHP, Heat Pipe, ammonia freezing, diffusion freeze-out, vapor-gas diffusion, ammonia frozen blockage*

## Nomenclature

**Subscripts**

| | |
|---|---|
| ad1 | adiabatic section 1 |
| ad2 | adiabatic section 2 |
| ev | evaporator |
| lv | ammonia in liquid-vapor state |
| res | reservoir |
| ss | ammonia in sublimated-solid state |

**Mathematical model parameters**

| | |
|---|---|
| HTC | heat transfer coefficient |
| $l_{GFL}$ | gas-vapor front location in the condenser |
| $n_{ad2}$ | number of moles in the adiabatic section 2 |
| $n_{cond}$ | number of moles in the condenser |
| $n_{res}$ | number of moles in the reservoir |

| | | |
|---|---|---|
| $n_{tot}$ | | total number of moles in the VCHP |
| $P_{g,ad2}$ | | NCG partial pressure in the adiabatic section 2 |
| $P_{g,cond}$ | | NCG partial pressure in the condenser |
| $P_{g,res}$ | | NCG partial pressure in the reservoir |
| $P_{v,\#}$ | | WF partial pressure in the NCG covered volume |
| $P_{v,ad1}$ | | WF pressure in the adiabatic section 1 |
| $\dot{Q}_{act}$ | | heat dissipation rate and input power |
| $\phi_{HP}$ | | VCHP diameter |
| $\mathcal{P}$ | | perimeter of the condenser wall |
| $R$ | | ideal gas constant |
| $S_{env}$ | | envelope surface area |
| $T_{ad1}$ | | temperature in the adiabatic section 1 |
| $T_{ad2}$ | | temperature in the adiabatic section 2 |
| $T_{cond}$ | | average temperature in the condenser |
| $T_{cond,a}$ | | temperature in the active part of the condenser |
| $T_{cond,i}$ | | temperature in the inactive part of the condenser |
| $T_{CS}$ | | cold source temperature |
| $T^{j}_{cond,i}$ | | local temperature in the inactive part of the condenser |
| $T_{ev}$ | | temperature in the evaporator |
| $T_{res}$ | | temperature in the reservoir |
| $T_{z_j}$ | | temperature distribution along the VCHP axis |
| $V_{ad2}$ | | volume of the adiabatic section 2 |
| $V_{cond,i}$ | | volume of the inactive portion of the condenser |
| $V_{res}$ | | volume of the reservoir |
| $z_j$ | | axial position inside the condenser |
| $\zeta$ | | computational routine tolerance |

**Acronyms**

| | |
|---|---|
| CCHP | Constant Conductance Heat Pipe |
| FFA | Flat Front Approach |
| GFL | Gas Front Location |
| NCG | Non-Condensable Gas |
| MFFA | Modified Flat Front Approach |
| VCHP | Variable Conductance Heat Pipe |
| WF | Working Fluid |

# 1 INTRODUCTION

A heat pipe is a device with a very high thermal conductivity capable of transporting significant amounts of heat over relatively short distances. Thanks to this characteristic, heat pipes are considered among the most efficient passive heat transfer devices [1–3]. Many types of heat pipes have been fabricated and designed to exploit their heat conducting properties in a wide range of applications. Examples include solar thermal and geothermal applications with an increasing popularity in automotive, nuclear and spacecraft industry, working from ultra-low to very-high temperature conditions [4–6]. The most standard type is a Constant Conductance Heat Pipe (CCHP), consisting of an external rod-shaped envelope and an inner capillary structure that covers the internal wall and is filled with the working fluid (WF) at saturation pressure [7,8]. Whenever the ends of the heat pipe are exposed to different temperatures, the working fluid evaporates at the hot end (evaporator) and the vapor, so generated, moves towards the cold end (condenser), dissipating the heat while changing back to liquid state. The condensed working fluid is then captured by the wick and transported by capillary forces to the evaporator [9–11]. This process is repeated in a loop. Other heat pipes commonly used in the industry are Variable Conductance Heat Pipes (VCHP hereinafter) [12], Loop Heat Pipes (LHP) [13,14], Capillary Pumped Loops (CPL) [15,16], Pulsating Heat Pipes (PHP) [17–19] and Micro Heat Pipes (MHP) [20–22].

A VCHP is a standard heat pipe with an additional reservoir of Non-Condensable Gas (NCG) attached to the end of the condenser [23]. The reservoir temperature can be either actively or passively feedback-controlled, regulating the heat dissipation in the condenser by moving the gas-vapor front position (usually referred to as Gas Front Location, GFL), changing thus the effective active length of the condenser [24]. This characteristic is very important, especially in spacecraft applications, where maintaining the temperature of the cooled devices as steady as possible is crucial [25]. In such applications, working fluids with low freezing temperatures and high range of operating temperatures are chosen, e.g. ethanol (-123°C to 241°C), propane (-188°C to 97°C) and ammonia (-78°C to 130°C) [1]. Ammonia in particular is widely used because of its high thermal capacity [26–30].

The portion of the VCHP filled with NCG represents a diffusion barrier to the flowing vapor. Consequently, the zone of the heat pipe near the gas-vapor front is characterized by the occurrence of binary mass diffusion between the vapor and the gas. Such process is usually neglected in the modeling stage of the design process, based on the assumption that the time scale of diffusion is orders of magnitude larger than the time scale of other processes like vapor convection, vapor-liquid phase change and heat dissipation. However, diffusion has been shown to play a very important role in certain scenarios [31,32], particularly when the VCHP works below freezing conditions [33–35], like those considered in the present study (ammonia/stainless-steel VCHP). When these heat pipes undergo freezing, experimental observations have shown that the decrease in the vapor concentration and the corresponding increase in the gas concentration occur smoothly over a non-negligible length of the pipe [36]. Taking into account the diffused nature of the gas-vapor interface instead of the standard flat-interface approach, Marcus et al. [36] were able to develop an analytical formulation for the temperature distribution.

Freezing can alter the normal start-up in ammonia heat pipes, possibly leading in extreme cases to permanent damage [37]. In CCHPs, for instance, the condenser core may become filled with frozen working fluid, resulting in dry-out of the evaporator [37,38]. This makes the start-up very complicated. To prevent dry-out in the evaporator, passive-controlled VCHPs are also designed to help in the start-up from a frozen state [37], in a way that the vapor pressure within the heat pipe decreases when the cold source temperature reaches freezing conditions. The NCG expands to maintain pressure equilibrium at the gas-vapor interface, thus reducing the active condenser length in the process. This confines the vapor to a small active portion, which is maintained at a higher temperature than an equivalent heat pipe with the same input heat power level. Because of its practical importance, the freezing process has been investigated in a number of works. Antoniuk et al. [35] studied the depletion of the working fluid in VCHPs, caused by the exposition of the condenser section to a series of freeze/thaw cycles. They were able to identify three types of freeze-out: i) *Suction freeze-out*, ii) *freezing blowby* and iii) *diffusion freeze-out*. Suction freeze-out occurs when the working fluid freezes close to the evaporator end of the VCHP, causing a mass depletion of the working fluid in the evaporator. Freezing blowby occurs when a frozen block is formed between the active condenser and the reservoir, thus creating a pressure difference between these two zones of the heat pipe. When the frozen block melts partially, the higher pressure in the active part pushes the working fluid from the evaporator towards the reservoir, resulting in a depletion of the working fluid in the active part. Detailed experimental investigation of freezing blowby in VCHP was carried out also in [35,39,40]. Finally, diffusion freeze-out is produced by the migration of vapor molecules of the working fluid from the active region near the gas-vapor front towards the inactive part of the condenser, where vapor condenses and freezes, triggering the working fluid depletion in the active part of the VCHP. Edwards et al. [41] carried out experiments on the diffusion freeze-out of water/copper VCHPs, focusing on the possible formation of a solid frozen block diffused through the gas-vapor front into the gas blocked part of the condenser. Similarly, Ellis et al. [38] studied the performance and the start-up of a titanium/water VCHP after extended periods of freezing, while Ochterbeck et al. [33] were able to visualize the formation of ice blockage bridging radially the inner part of the VCHP near the zone of the gas-vapor front in the active section of the condenser. However, to this date, not much has been researched about the diffusion freeze-out blockage formation in VCHPs when a working fluid other than water (e.g. ammonia, where the volume expansion before freezing does not occur [42]) is used. In addition, none of the above-mentioned works has addressed the formation of a frozen working fluid block by diffusion freeze-out in ammonia/stainless-steel VCHPs. Consequently, little is known about the possibility of a thin frozen block formation in the inactive part of the VCHP at ammonia freezing point conditions, about the time scale that characterizes the process and about the possible correction practices. The purpose of this paper is to fill these knowledge gaps and it is motivated by the widespread use of ammonia/stainless-steel VCHPs in a number of different spacecraft. These are inevitably exposed to ultra-low temperatures below -80°C over long periods of time [37,38,43]. The present study also aims at demonstrating the importance of diffusion in such subfreezing conditions, providing a non-visual procedure to detect a frozen blockage formation and offering an estimate of the exposure time necessary for the block formation. This information will be relevant for the industrial design and management of thermal systems in order to scientifically justify the actions currently taken as preventive practices to avoid any malfunctioning in the spacecraft. Since these actions require additional weight and added energy consumption to the spacecraft, weight being a critical design parameter and energy being a limited resource, stating whether they are (or are not) necessary, could greatly impact the cost of the VCHP operation in space.

The paper is structured as follows: The mathematical model, the measurement methods and the preliminary operational characteristics of the VCHP are provided in Sec. 2. The experimental set-up is described in Sec. 3, and the experimental outcomes are shown and discussed in Sec. 4. Finally, the main conclusions are presented in Sec. 5.

## 2 METHODS

### 2.1 VCHP CHARACTERISTICS

The VCHP used in this experiment (a spare sample of the ATV – Automated Transfer Vehicle – program) has five parts, as shown in Figure 1. The first part is the evaporator, which is followed by a first adiabatic section (referred to as Adiabatic 1 hereinafter) and by the condenser, which is connected to a second adiabatic section (referred to as Adiabatic 2 hereinafter). The last part is the reservoir. This VCHP was designed and built by Euro Heat Pipes (Belgium) for aerospace applications where the heat pipe is exposed to very-low temperatures for extended periods. For this reason, the device is made of stainless steel, with a wick mesh on the inner pipe wall that extends all the way to the reservoir, which can be thermally controlled (i.e. the reservoir temperature $T_{res}$ can be varied using a heater mounted on the external surface). The working fluid is ammonia and the used NCG is nitrogen.

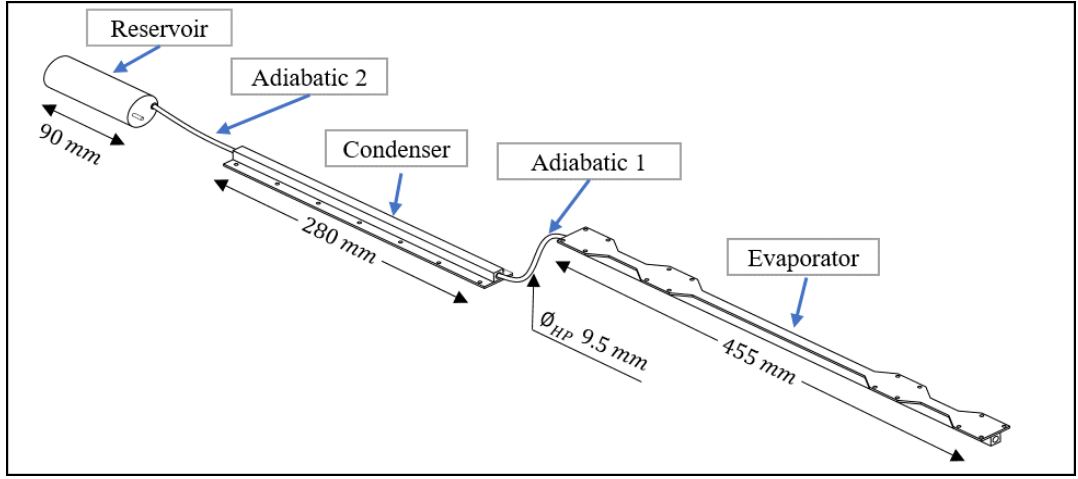
*Figure 1 Isometric view of the Variable Conductance Heat Pipe examined in the present study.*

## 2.2 MODIFIED FLAT FRONT APPROACH

The Flat Front Approach (FFA) is a simplified model of a VCHP. It is widely used at the stage of the preliminary design, because it gives clear insights into the VCHP operational characteristics and is easy to implement. FFA assumes a sharp interface between the gas and vapor [44]. However, under certain conditions (e.g. subfreezing temperatures) a diffused description of the gas-vapor interface is required in order to describe diffusion and freezing operational parameters, especially in the inactive part of the condenser. We therefore define a modified Flat Front Approach (MFFA), which is an extended version of the Flat Front Approach, to consider the effects of the axial conduction in the condenser wall. This allows a more detailed description of the temperatures in the inactive portion of the condenser, which is then treated as an axially conductive fin. Then a fin equation then can be derived to approximate the temperature profile in this zone.

To predict the temperature profile of the VCHP and to calculate the GFL, assuming steady state conditions and uniform total pressure throughout the VCHP, the MFFA is based on the NCG distribution affected by its expansion and compression inside of the heat pipe. The main step of this approach is computing the total number of moles $n$ of NCG present in the relevant sections of the VCHP based on the ideal gas equation:

$$n = \frac{PV}{RT}, \qquad (1)$$

where $P$ is the gas pressure, $V$ is the volume occupied by the gas, $R$ is the ideal gas constant with a value of (8.31 m³Pa mol⁻¹ K⁻¹) and $T$ is the gas temperature. NCG occupies not only the reservoir, but also the nearby the Adiabatic 2 and a part of the condenser. Therefore, the total number of moles is:

$$n_{tot} = n_{res} + n_{ad2} + n_{cond}, \qquad (2)$$

where $n_{res}$, $n_{ad2}$ and $n_{cond}$ are the amounts of moles at each of the above-mentioned sections. Equation (2) can be rewritten as:

$$n_{tot} = \frac{P_{g,res} V_{res}}{RT_{res}} + \frac{P_{g,ad2} V_{ad2}}{RT_{ad2}} + \frac{P_{g,cond} V_{cond,i}}{RT_{cond,i}}, \qquad (3)$$

where $P_{g,res}$, $P_{g,ad2}$ and $P_{g,cond}$ represent the NCG partial pressures at the reservoir, at the Adiabatic 2 and at the condenser, respectively, whereas $V_{res}$, $V_{ad2}$, $V_{cond,i}$ represent the total volumes of the reservoir, the Adiabatic 2 and the inactive part of the condenser, respectively. Since all the zones inside the VCHP must be at saturation pressure (referred to as $P_{v,ad1}$ hereinafter), the partial pressures in equation (3) can be calculated considering the partial pressure of the WF in the sections filled with NCG (labelled as $P_{v,\#}$) as follows:

$$P_{g,\#} = P_{v,ad1} - P_{v,\#}. \tag{4}$$

In equation (4), $P_{v,\#}$ is obtained from known vapor-pressure correlations for ammonia [45]. We also define a local value of the temperature, labelled $T_{cond,i}^j$, with the subscript $i$ representing the inactive portion of the condenser, and the superscript $j$, representing the specific location within the inactive portion at which the temperature is calculated (in this work, we considered 20 uniformly-spaced locations). At this temperature, the partial pressures of the working fluid correspond to the saturation conditions, as our VCHP has a wicked reservoir.

To compute $T_{cond,i}^j$, we can exploit a conventional fin analysis (see Appendix B), since the WF condensation is dramatically reduced in the inactive region due to the presence of NCG, and the axial conduction dominates over the thermal diffusion [44]. This leads to the following equation:

$$T_{cond,i}^j = T_{CS} + (T_{cond,a} - T_{CS}) \exp\left[\sqrt{\frac{HTC \cdot \mathcal{P}}{kA_c}}(z_j - l_{GFL})\right], \tag{5}$$

where $T_{CS}$ is the cold source temperature, being the lowest temperature in the condenser (and in fact, in the entire heat pipe) that is produced by a cold source (also referred to as heat sink) placed around the condenser and constituted by a heat exchanger with liquid hydrogen as coolant; $T_{cond,a}$ [K] is the temperature in the active portion of the condenser, $HTC$ [W/m²/K] is the heat transfer coefficient (a measured value taken from previous VCHP tests), $\mathcal{P}$ [m] is the perimeter of the condenser wall (usually referred to as envelope) in contact with the cold source, $k$ [W/m/K] is the axial conductivity of the envelope material, $A_c$ [m²] is the cross sectional area of the envelope, $z_j$ is the axial position selected within the condenser and $l_{GFL}$ [m] is the length of the inactive portion of the condenser.

The temperature $T_{cond,a}$ can be computed using the following equation:

$$T_{cond,a} = T_{ad1} - \frac{\dot{Q}_{act}}{HTC \, S_{env}}, \tag{6}$$

where $T_{ad1}$ is the temperature in the Adiabatic 1 (which is almost equal to the temperature in the evaporator), $S_{env}$ [m²] is the VCHP's envelope surface area and $\dot{Q}_{act}$ [W] is the heat dissipation rate (due to the working fluid condensation) measured in the active part of the condenser. The heat dissipation rate in the active part of the condenser is assumed to have the same value as the input power in the evaporator, and both are represented by $\dot{Q}_{act}$.

The equations presented above are implemented as an iterative routine, the steps of which are provided in the flowchart of Figure 2. To start the routine, an initial guess value for $T_{ad1}$ must be assumed, which allows to obtain $P_{v,ad1}$ and subsequently to solve equation (4) and equation (6). In the first iteration, the value of $V_{cond,i}$ is computed using equation (3) and considering that the inactive part is uniform and equal to the cold source temperature, i.e. $T_{cond,i} = T_{CS}$. For the subsequent iterations, the modeled temperature calculated from the previous iteration, $T_{cond,i} = T_{cond,i}^j$, is used. Then, the gas-front location ($l_{GFL}$) can be calculated using $V_{cond,i} = A_{cond} * l_{GFL}$, where $A_{cond}$ is the cross-sectional area of the condenser. The parameters computed up to this step of the routine allow to evaluate the temperature profile in the inactive part of the condenser $T_{cond,i}^j$ using equation (5). The next step in the routine is to compute the number of moles in the current iteration, $n'_{tot}$, using equation (3). Subsequently, the condition $|(n_{tot}-$

$n'_{tot})/n_{tot}|<\zeta$, where $\zeta$ represents the tolerance, is evaluated. If this condition is not met, the iterative calculation continues, and two scenarios are possible: if $n'_{tot}$ is smaller than $n_{tot}$, then temperature in the Adiabatic 1 in the current iteration ($T^t_{ad1}$) is decreased by $dT$ to obtain the value of the next iteration ($T^{t+1}_{ad1}$); if $n'_{tot}$ is larger than the value of $n_{tot}$, then $T^t_{ad1}$ is increased by $dT$ to obtain $T^{t+1}_{ad1}$. In either case $T^{t+1}_{ad1}$ is used as an input (instead of the guessed value) in the subsequent iteration and the routine is run again. The iterative calculation continues until the condition $|(n_{tot}-n'_{tot})/n_{tot}|<\zeta$ is fulfilled.

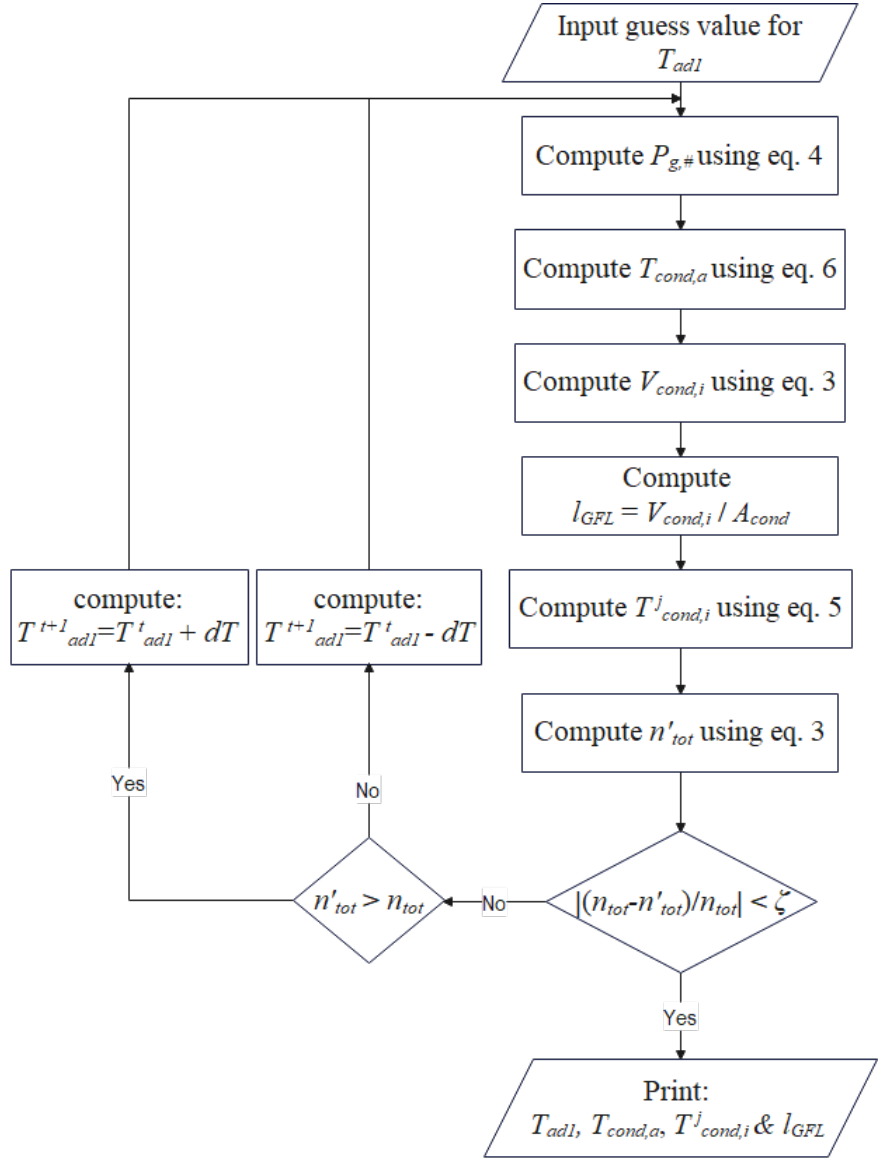

*Figure 2 MFFA computational routine flowchart: the aim is to compute the operational parameters of the VCHP such as $l_{GFL}$, $T^j_{cond,i}$, $T_{cond,a}$ and $T_{ad1}$.*

## 2.3 CONCEPTUALIZATION OF EXPERIMENTAL CONDITIONS

To understand the parameters involved in the diffusion freeze-out, we follow the procedure presented by Edwards et. al. [41], which starts by considering Fick's first law to determine the vapor diffusion flux into the inactive part of the VCHP:

$$J = -cD_P \frac{dx}{dz_j} = -cD_P \frac{dx}{dT_{z_j}} \cdot \frac{dT_{z_j}}{dz_j}, \tag{7}$$

where $J$ [mol s$^{-1}$ m$^{-2}$] is the vapor diffusion flux, $c$ [mol m$^{-3}$] is the molar density, $D_P$ [$Pa\ m^2\ s^{-1}$] is the diffusivity of gas-vapor and $x$ is the mole fraction of the vapor, the variable $z_j$ [m] represents the axial position of the condenser and $T_F$ is the temperature of the working fluid at the freezing point.

In the zone of the gas-vapor front, the axial conduction effect dominates over the axial diffusion. Therefore, the thermal conduction defines the axial temperature profile [41,44]. The temperature behavior for the portion of the VCHP containing NCG can be approximately described using a conventional fin equation (derivation shown in Appendix B):

$$T_{z_j} = T_{cs} + (T_F - T_{cs}) \exp\left[\sqrt{\frac{HTC \cdot \mathcal{P}}{kA_c}}(z_j - z_F)\right], \tag{8}$$

where the variable $T_{z_j}$ represents the local axial temperature in the condenser and $z_F$ is the axial position where freezing occurs. Taking the derivative of equation (8) with respect to $z_j$ and evaluating it at the position $z_F$, we get:

$$\frac{dT_{z_j}}{dz_j} = (T_F - T_{cs})\sqrt{\frac{HTC \cdot \mathcal{P}}{kA_c}}. \tag{9}$$

Using the Clausius-Clapeyron equation and considering the relation between the WF vapor partial pressure ($P_{v,\#}$) and the vapor mole fraction ($x = P_{v,\#}/P$), we get:

$$x = \exp\left[\frac{-\lambda}{RT_{ev}}\left(\frac{T_{ev}}{T_{z_j}} - 1\right)\right], \tag{10}$$

where $T_{ev}$ is the temperature in the evaporator, $\lambda$ [J/mol] is the enthalpy of vaporization and $R$ is the universal constant for ideal gases. Then we take the derivative of equation (10) with respect to $T_{z_j}$ and we evaluate it at the freezing temperature $T_F$:

$$\frac{dx}{dT_{z_j}} = \frac{\lambda}{RT_F^2} \exp\left[\frac{-\lambda}{RT_{ev}}\left(\frac{T_{ev}}{T_F} - 1\right)\right]. \tag{11}$$

Finally, equations (9) and (11) are incorporated into equation (7):

$$J = \frac{-c\,D_P\,\lambda(T_F - T_{cs})}{RT_F^2} \exp\left[\frac{-\lambda}{RT_F}\left(1 - \frac{T_F}{T_{ev}}\right)\right] \sqrt{\frac{HTC \cdot \mathcal{P}}{kA_c}}. \tag{12}$$

Equation (12) incorporates all the main parameters acting for and against, which can favor or prevent the vapor diffusion through the gas-vapor front. To enhance the diffusion rate and thus increase the chances to form a WF frozen block, special attention must be paid to the terms that can be controlled in equation (12), considering the real set-up of the experiment. Two parameters that can be tuned independently are: i) $T_{ev}$, which can be modified by changing the heat power input and/or the reservoir temperature $T_{res}$; ii) $T_{cs}$, which can be varied directly by the cooling system. Increasing the difference between these two temperatures would strengthen the vapor flux diffusion towards the inactive part of the condenser.

One of the main concerns of the present work is to find a way to determine if a frozen ammonia blockage has been formed in the inactive part of the condenser. No direct visual access to the VCHP was possible during the experiments. Therefore, we devised an indirect method to detect the presence of a blockage. The working principle is based on the motion of the gas-vapor front and the subsequent variation of temperature in the Adiabatic 1, induced by a change of temperature in the reservoir.

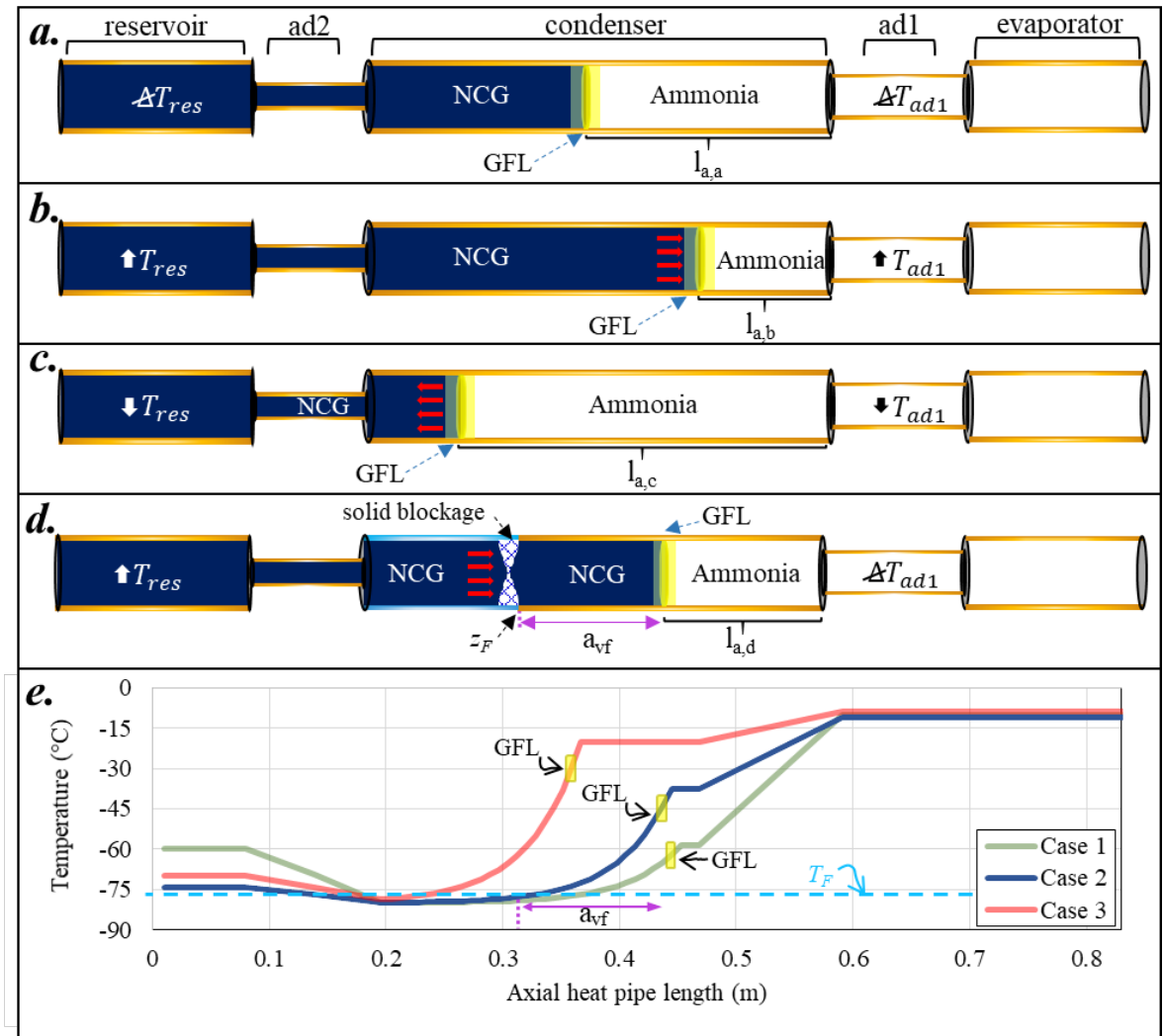

*Figure 3 **a.** Schematic representation of a VCHP working at steady state. **b.** Schematic explanation of GFL motion and $T_{ad1}$ change, caused by $T_{res}$ increment. **c.** Schematic explanation of GFL motion and $T_{ad1}$ change, caused by $T_{res}$ decrement. **d.** Schematic ilustration of a VCHP with a blockage formed in the condenser. While $T_{res}$ is increased, the GFL and $T_{ad1}$ remain unchanged. **e.** Preliminary VCHP temperature profiles generated using MFFA to better understand the operational characteristics of the VCHP in different scenarios.*

We can illustrate this idea by means of Figure 3. Panel **a.** shows a VCHP working at steady state with an active heat dissipation length $l_{a,a}$. Panel **b.** illustrates the case where the reservoir temperature is increased. This increment leads to the expansion of the NCG into the condenser, moving the gas-vapor front towards the evaporator and thus shortening the active heat dissipation length to $l_{a,b}$. As a consequence, the temperature in the Adiabatic 1 rises. The opposite case is shown in panel **c.**, where the temperature in the reservoir is decreased: this leads to the heat dissipation length $l_{a,c}$ extension, decreasing thus the temperature in the Adiabatic 1. From the cases discussed up to this point, we infer that, as long as the gas-vapor front is free to move, temperatures in the reservoir and the Adiabatic 1 are correlated.

Figure 3.d depicts the case where an ammonia frozen blockage has been formed close to the freezing point, dividing the inactive part of the condenser into two parts. As in the previous case, the temperature in the reservoir is increased, but the NCG cannot expand freely towards the evaporator anymore, because it is confined by the blockage. This means that the active dissipation length $l_{a,d}$ does not change, and by extension, the temperature in the Adiabatic 1 does not vary at all.

The principle described above explains how we can experimentally test the temperature correlation between the reservoir and the Adiabatic 1 during the subfreezing experiments (detailed later in Sec.4). We establish that a blockage has been successfully detected when the temperature correlation is broken (case in Figure 3.d).

If a thin frozen blockage is formed in the inactive portion of the condenser, then it will appear in the vicinity zone of the freezing point ($z_F$). The modified Flat Front Approach can help us to generate the VCHP temperature profile for different working parameters and to locate the freezing point location (given by the intersection of the temperature profile and the horizontal light-blue dashed line shown in Figure 3.e). Based on the location of $z_F$ we can foresee three general scenarios, referred to as cases in Figure 3.e (with a temperature gradient between the evaporator and the cold source of at least 70°C). In Case 1, the points $z_F$ and GFL are very close. This can bring on some problems for a stable blockage formation in the inactive section, like partial melting or trapped pockets of hot ammonia vapor interacting directly with the blockage (caused by accidental vapor fluctuations during the experiments). Case 3, on the other hand, presents the points $z_F$ and GFL far from each other, a small portion of the inactive part is exposed to subfreezing temperatures and $z_F$ is close to the Adiabatic 2. However, since the path of diffused ammonia from GFL to $z_F$ is long, some amount of it would still be able to condense and return to the active part, making the blockage formation slower and harder.

Case 2 shows a good compromise of the distance from $z_F$ to GFL (denoted as $a_{vf}$ in Figure 3.e), i.e. the formation of a blockage will be isolated and undisturbed, diffusion being the only driving mechanism which carries ammonia to $z_F$ (Figure 3.d). It also shows an important portion of the inactive part at subfreezing conditions and a smooth temperature profile in the transition from $z_F$ to GFL. Therefore, Case 2 represents the optimal scenario for the experimental setup.

## 3 EXPERIMENTAL SETUP

To investigate the possible formation of a frozen block of ammonia (radially covering the HP core) thermally diffused inside the inactive section, freezing tests were performed to a VCHP at EHP facilities. The experimental setup used for the tests is described in this section.

The test bench consisted of a horizontal board 3 m long and 1 m wide, equipped with two liquid nitrogen feeding pipes, 42 J-type thermocouples (TC), a precise tilt control system, a heat power input and a control system. To thermally insulate the VCHP from the laboratory environment, polystyrene beads were used. The VCHP was instrumented with 31 TCs, placed as shown in Figure 4. The TCs were arranged in the different sections as follows:

- o  3 TCs in the evaporator,
- o  1 TC in the Adiabatic 1 and 1 TC in the Adiabatic 2,
- o  20 TCs in the condenser,
- o  3 TCs for the reservoir,
- o  3 security TCs.

20 TCs were placed in the upper and lower side of the condenser (distributed in 10 positions, i.e. 2 TCs per position), which ensures a more accurate measurement of the temperature profile in this section. The TCs in the evaporator, the reservoir and the adiabatic parts were installed only in the upper part of the heat pipe, as shown in Figure 4. For better reference of the position of the TCs, two different horizontal axes $z_{O1}$ (positive values to the right) and $z_{O2}$ (positive values to the left) were defined, with origin O1 and O2, respectively. O1 coincides with the free end of the reservoir and O2 with the free end of the evaporator. Table 1 provides the location of the TCs using $z_{O1}$ and $z_{O2}$. Security TCs were placed in the evaporator (which is the hottest part of the VCHP) to shut off the heater if temperatures above 60°C are reached, in the reservoir and in the condenser to turn off the liquid nitrogen cooling system if temperatures below -95°C are reached.

The TCs measurements were recorded every 5 seconds (with an accuracy of 0.2°C verified through calibration of the whole measurement chain from the TC to the computer) to have a detailed time evolution of the monitored temperatures. The measurement consisted of 4 stages, reported in Table 2. First, a health check was performed to ensure that the VCHP is not malfunctioning, followed by a temperature profile stabilization stage and a subsequent check to verify that no blockage was generated. Then, the freezing test was carried out and finally, another blockage check was performed; if a frozen ammonia blockage is detected at this stage, the condenser is heated up to remove the blockage.

Upon equipping the VCHP with the TCs, three Dale RH-50 50 W power resistors were installed on the hot plate in the evaporator, and cooling devices (heat exchangers using the constant flow of liquid nitrogen) were placed in the condenser and in the reservoir. The instrumented VCHP was mounted on the test bench, as shown in Figure 5. The TCs and the power resistors were connected to the Data acquisition system. The test bench and the VCHP were then covered with polystyrene beads and finally, the cooling system was turned on to begin with the freezing tests.

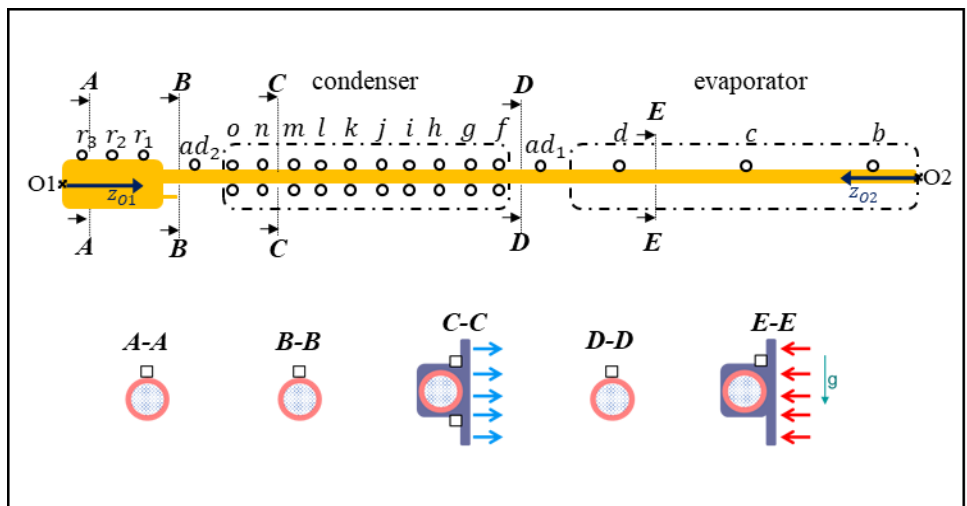

*Figure 4 Arrangement of the thermocouples (TCs) along the VHCP. The diagram at the bottom shows the cut-view of the TCs position in each part of the heat pipe. The arrows in the views C-C and E-E indicate the direction of the heat flux.*

*Table 1 Axial position of the thermocouples (TCs) along the heat pipe. Subscripts "up" and "low" refer to the TCs installed in the upper and lower part of the heat pipe, respectively.*

| TC label | Origin | Position (mm) | TC label | Origin | Position (mm) |
|---|---|---|---|---|---|
| $b$ | $z_{O2}$ | 150 | $k_{up}, k_{low}$ | $z_{O1}$ | 305 |
| $c$ | $z_{O2}$ | 245 | $l_{up}, l_{low}$ | $z_{O1}$ | 285 |
| $d$ | $z_{O2}$ | 340 | $m_{up}, m_{low}$ | $z_{O1}$ | 265 |
| $ad_1$ | $z_{O2}$ | 533 | $n_{up}, n_{low}$ | $z_{O1}$ | 245 |
| $f_{up}, f_{low}$ | $z_{O1}$ | 445 | $o_{up}, o_{low}$ | $z_{O1}$ | 205 |
| $g_{up}, g_{low}$ | $z_{O1}$ | 405 | $ad_2$ | $z_{O1}$ | 137.5 |
| $h_{up}, h_{low}$ | $z_{O1}$ | 365 | $r_1$ | $z_{O1}$ | 80 |
| $i_{up}, i_{low}$ | $z_{O1}$ | 345 | $r_2$ | $z_{O1}$ | 45 |
| $j_{up}, j_{low}$ | $z_{O1}$ | 325 | $r_3$ | $z_{O1}$ | 10 |

*Table 2 Sequence of the measurement stages.*

| Stage | $T_{res}$ [°C] | $T_{ad1}$ [°C] | Power (evaporator) [W] | Time span | Notes |
|---|---|---|---|---|---|
| Health check | -30 | 4.3 | 25 | Stable or 1 hour | Cold source at 1.3°C |
| Stabilization & Blockage check | Variable | Variable | 70 | 3 hours | Cold source at -80°C |
| Freezing experiment on VCHP | -74 | -10.7 | 70 | 211 hours | Cold source at -80°C |
| Blockage check | Variable | Variable | 70 | 3 hours | Cold source at -80°C |
| Blockage removal | -74 | Variable | 70 | 2 hours | Variable $T_{cs}$ |

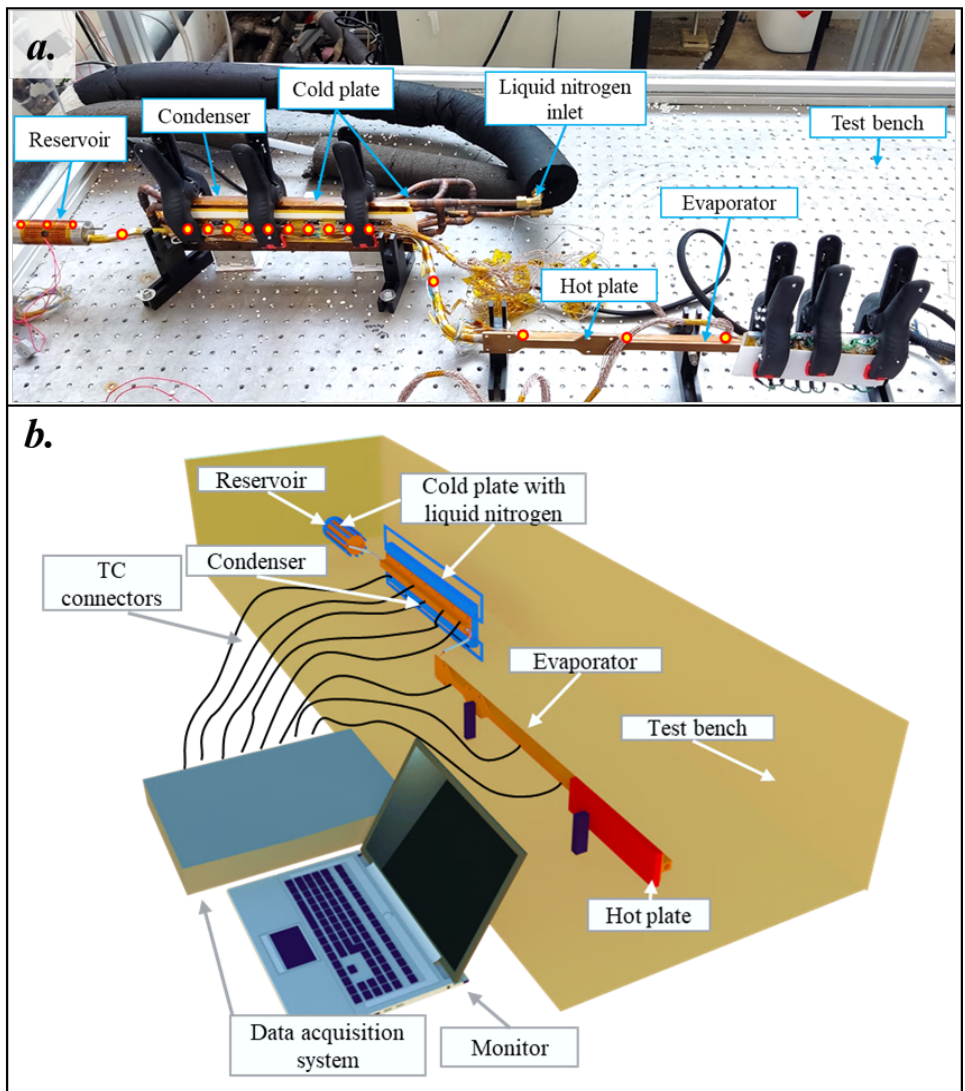

*Figure 5 **a**. Photo of the experimental setup mounted on the test bench, the red and yellow circles show the TCs poisition. **b**. Visualization of the experimental setup mounted on the test bench.*

Unless otherwise specified, the allowable tolerances and the experimental conditions shall be as indicated in the Table 3, where the steady-state condition is a scenario with low temperature fluctuations.

*Table 3 Experimental conditions and allowed tolerances.*

| Condition | Value |
|---|---|
| Range of temperatures for the experiments | -95°C <T<5°C |
| Power input tolerance | < 2.5% |
| Thermocouple precision | ± 0.2°C |
| Humidity | 60% ± 20% RH |
| Ambient temperature | 20°C ± 5°C |
| Data acquisition frequency | 1/5 s$^{-1}$ |
| Steady-state condition | ΔT < 0.5°C for 2 hours |

## 4 RESULTS AND DISCUSSION

In this section, we present the VCHP thermal characteristics achieved by experimentation and by its corresponding model using MFFA. The outcomes of each of the measurement stages (presented in Table 2) are then provided and further discussed.

### 4.1 TEMPERATURE PROFILE AND MODEL VERIFICATION

The measured temperature profile shown in Figure 6.a (performed at steady-state condition) matches the operational characteristics of the selected case in Section 2, i.e. a portion of the inactive part reaches temperatures below the ammonia freezing point ($T_F$ = -77.78°C), the freezing point location ($z_F$) is far from the gas-vapor front (GFL) and the temperature profile presents a smooth transition with respect to the freezing point.

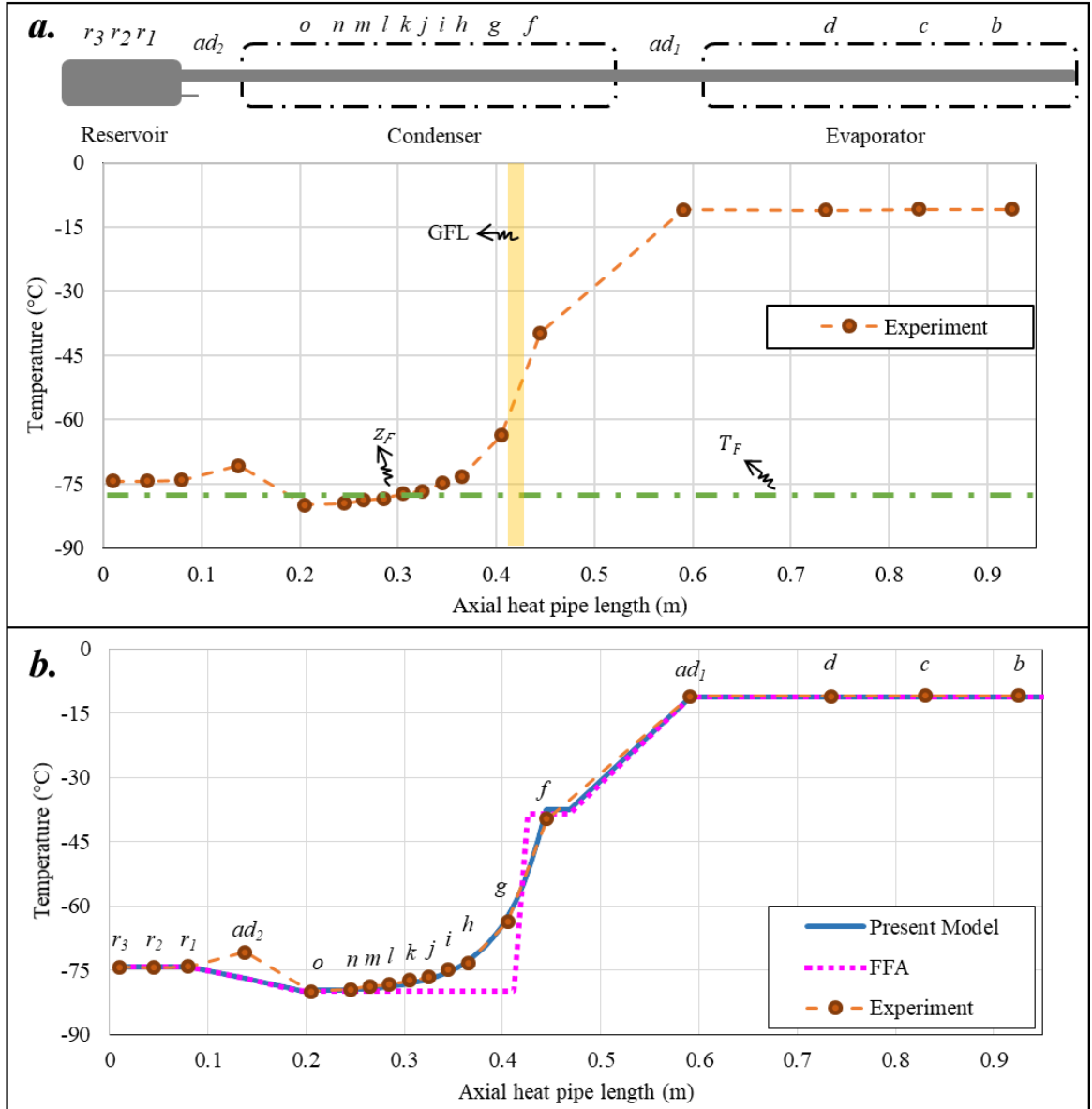

Figure 6 *a.* Schematic layout of the thermo-couples position in the VCHP and the plot of its measured temperature profile. *b.* Present model (MFFA), FFA and experimental temperature profile comparison, where the GFL lies between points " *f* " and " *g* ".

To verify the temperature profiles resulting from the MFFA, a total of nine experimental case studies were carried out under different thermal and heat power conditions. The temperature profiles computed by the mathematical model were found to be in good agreement with the experimental measurements for all the cases. Eight of them are reported in Appendix A (see Figure A 1), while Figure 6.b shows the case selected for the freezing experimental tests. The gas-front location obtained using the MFFA is $z_{O1} \approx 0.42$ m, which falls between points "*f*" and "*g*".

The results presented in Figure 6 were obtained using the parameters indicated in Table 4.

*Table 4 MFFA and VCHP operational parameters.*

| Parameter | Value | Unit |
|---|---|---|
| $dT$ | 1x10$^{-2}$ | [°C] |
| $HTC$ | 2033 | [W/(m$^2$K)] |
| $n_{tot}$ | 1.46x10$^{-2}$ | [mol] |
| $\dot{Q}_{act}$ | 70 | [W] |
| $\phi_{HP}$ | 9.5x10⁻³ | [m] |
| $k$ | 25 | [W/(m K)] |
| $T_{res}$ | -74.30 | [°C] |
| $T_{CS}$ | -80 | [°C] |
| $V_{res}$ | 6.95x10⁻⁶ | [m³] |
| $\zeta$ | 1x10⁻² | - |

Figure 6.b also includes the resulting temperature profile using a conventional FFA (magenta dashed curve). This is characterized by a step-like shape, where the transition from the temperatures in the active section to the temperatures in the inactive section is considered as a sharp interface, which is well represented by the gas front location. . However, this assumption does not allow to model the temperature evolution along the inactive portion of the condenser. In order to include this information, we used a modified Flat Front Approach (blue solid curve in Figure 6.b), which enabled us to understand the operational characteristics of the VCHP and to build a general concept for the experimental setup, as described in Sec. 2.3.

## 4.2 HEALTH CHECK

The health check is a procedure by which the VCHP is tested and compared with older operative data to verify that the device is suitably calibrated and also works properly. The check was performed with the following input data for a fully open VCHP taken from the VCHP fabrication data sheet (available upon request to EHP): $T_{res}$=-30°C, $T_{CS}$=1.3°C and $T_{sat}$= 4.3°C, for $\dot{Q}_{act}$=25 W.

Figure 7 shows the experimental temperature profile result (at steady-state condition), where the condenser temperature is constant at an average value of $T_{cond}$= 3.04 °C. The mathematical model calculates the temperature profile (plotted alongside the experimental one) and determines the GFL to be at $z_{O1} \approx 0.2$ m (between point "o" and point "$ad_2$"), i.e. the NCG is retreated towards the reservoir and the condenser is full of ammonia vapor. To further verify that the VCHP is fully open, equation (6) is used, together with the resulting temperatures, the condenser dissipation area ($S_{env}$= 0.0094 m²) and the data in Table 4, to determine the heat rate dissipation. The calculated value is 24.7 W, which is similar to the input power, proving that the active dissipation length is equal to the condenser length, and by extension, that the VCHP is calibrated.

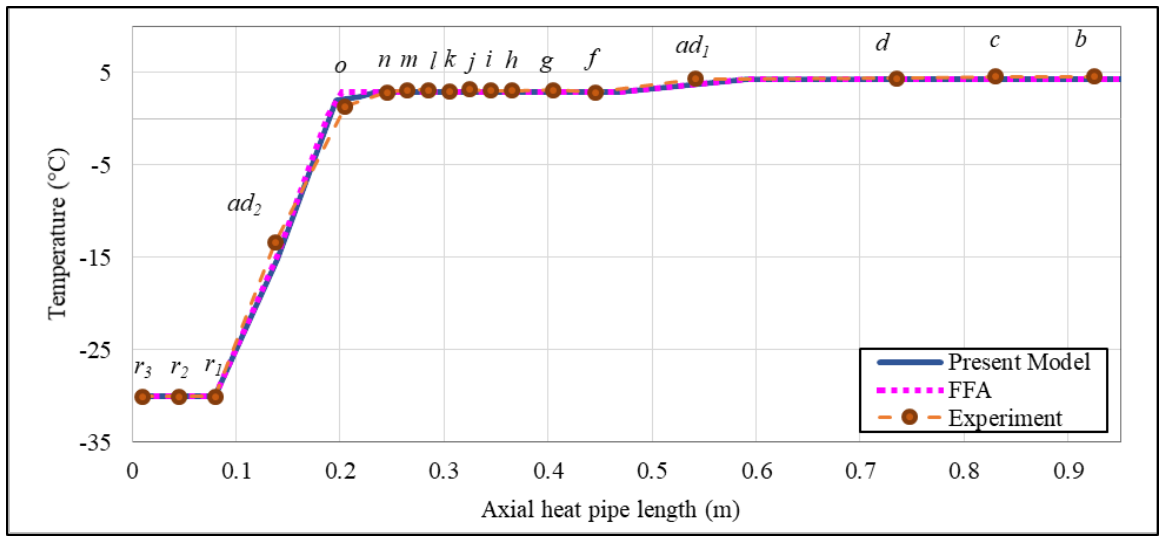

*Figure 7 Health check experimental and computed temperature profile.*

## 4.3 STABILIZATION

During the experiments, the stabilization procedure was employed twice. This process serves the purpose of bringing the temperature distribution along the VCHP to the steady-state temperature profile shown in Figure 6.a using the operational parameters described in Table 4. The first stabilization was performed after the health check. The persistence of a stable temperature profile was then monitored for as long as 20 hours: during this time span, no blockage was detected. The second stabilization process was applied after the blockage check and before the subfreezing experiment, Table 5 reports the temperatures measured by the TCs at steady-state conditions (time span of 2 hours). It also summarizes the relevant measurement uncertainty parameters, evidencing the – rather limited – spread of temperature fluctuations around the mean.

*Table 5 Local mean values of the stabilized temperature profile along the VCHP, along with the standard deviation (SD), the standard error of mean (SEM), the upper and lower limit of the confidence interval 95%, (CI 95% u and CI 95% l).*

| Point | Mean (°C) | SD (°C) | SEM (°C) | CI 95% u (°C) | CI 95% l (°C) | Point | Mean (°C) | SD (°C) | SEM (°C) | CI 95% u (°C) | CI 95% l (°C) |
|---|---|---|---|---|---|---|---|---|---|---|---|
| $T_{res}$ | -74.30 | 0.28 | 0.0626 | -74.1772 | -74.4227 | $T_j$ | -76.70 | 0.16 | 0.0357 | -76.6298 | -76.7701 |
| $T_{ad2}$ | -70.80 | 0.25 | 0.0559 | -70.6904 | -70.9095 | $T_i$ | -74.83 | 0.18 | 0.0402 | -74.7511 | -74.9088 |
| $T_o$ | -80.00 | 0.20 | 0.0447 | -79.9123 | -80.0876 | $T_h$ | -73.35 | 0.17 | 0.0380 | -73.2754 | -73.4245 |
| $T_n$ | -79.75 | 0.21 | 0.0469 | -79.4579 | -79.6420 | $T_g$ | -63.74 | 0.16 | 0.0357 | -63.6698 | -63.8101 |
| $T_m$ | -78.78 | 0.19 | 0.0424 | -78.6967 | -78.8632 | $T_f$ | -39.74 | 0.16 | 0.0357 | -39.6698 | -39.8101 |
| $T_l$ | -78.40 | 0.18 | 0.0402 | -78.3211 | -78.4788 | $T_{ad1}$ | -11.00 | 0.08 | 0.0178 | -10.9649 | -11.0350 |
| $T_k$ | -77.40 | 0.16 | 0.0357 | -77.3298 | -77.4701 | $T_{ev}$ | -11.06 | 0.08 | 0.0178 | -11.0249 | -11.0950 |

We remark that the occurrence of a cross-over of the ammonia freezing point ($T_F$ = -77.78°C) is still between points "*l*" and "*k*" ($T_l$ and $T_k$ being right below and right above $T_F$, respectively), and we recall that the GFL (obtained using MFFA) is between points "*g*" and "*f*".

## 4.4 BLOCKAGE CHECK

Starting from the stabilized temperature profile, the presence of a blockage was checked during the initial (short-term) exposure to freezing conditions. The check is carried out by perturbing the temperature $T_{res}$ and measuring the corresponding change of $T_{ad1}$, as shown for example in Figure 8. In this case, $T_{res}$ was changed to produce a smooth variation characterized by alternated reductions and increases. In particular, $T_{res}$ was first cooled down from -74.3°C to -89.0°C. $T_{ad1}$ reduces proportionally at the same rate from -10.7°C to -13.0°C, meaning that a temperature reduction of 6.4°C in the reservoir corresponds to a temperature reduction of 1°C in the Adiabatic 1. Two subsequent and milder increments of temperature (from 2.3 to 2.4 hours, and later from 2.5 to 2.6 hours) were performed for $T_{res}$, producing a similar effect in the Adiabatic 1. The strong correlation characterizing the instantaneous behavior of $T_{res}$ and $T_{ad1}$ indicates that no blockage has occurred, as in fact, the presence of a blockage would decouple $T_{ad1}$ from $T_{res}$.

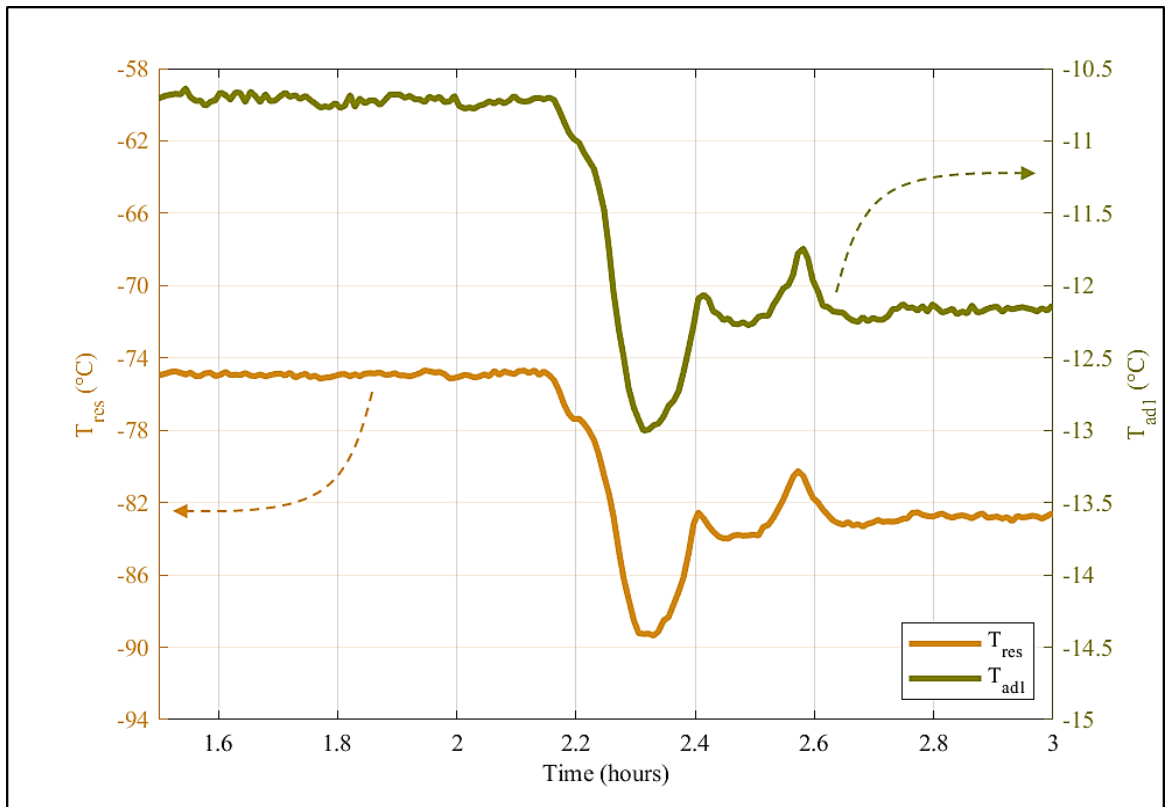

*Figure 8 Blockage check after short-term exposure to freezing conditions and smooth temperature variation in the reservoir. The arrows indicate the vertical axis where the range of temperatures corresponding to each curve is provided.*

The occurrence of a blockage after a short-term exposure to freezing conditions does not seem to be related to the smoothness of the temperature variation applied in the reservoir. Indeed, even when $T_{res}$ is modified abruptly, the profiles of $T_{res}$ and $T_{ad1}$ remain correlated, as shown in Figure 9. It should be emphasized that the blockage checks described above were made at rather early stages of the exposure to freezing conditions and over a time span of about three hours. Such time scale is too short to be able to observe the formation of a blockage formed by diffused frozen ammonia. Diffusion-induced blockages are indeed expected to form over a much longer time scale, which can be estimated as $t_{diff} = \mathcal{L}^2/\mathcal{D}$, where $\mathcal{L} \cong O(10^{-2})$ m represents the characteristic distance traversed by diffused molecules and $\mathcal{D}$ is the diffusion coefficient of liquid ammonia in a gas medium. Knowing that the self-diffusion coefficient and the diffusion coefficient of vapor in a gas medium have the same order of magnitude [47,48], we take the value of the self-diffusion coefficient for ammonia at -73.15°C reported in [49,50], and use the same order of magnitude, which is $\mathcal{D} \cong O(10^{-9})$ m$^2$s$^{-1}$. The estimated diffusion time scale is then in the order of magnitude of $t_{diff} \cong O(10^5)$ s (a time range between 28 and 55 hours).

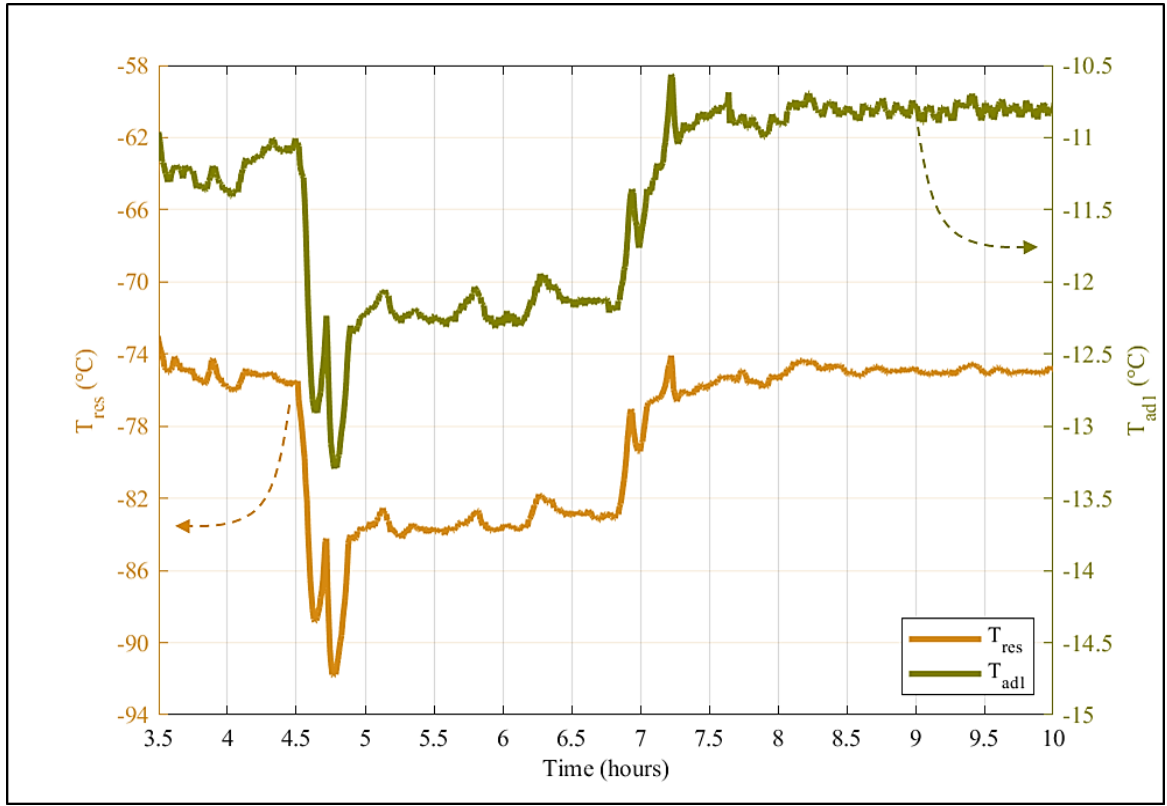

*Figure 9 Blockage check after short-term exposure to freezing conditions and abrupt temperature variation in the reservoir. Arrows have the same meaning as in Figure 8.*

### 4.5 LONG-PERIOD EXPOSURE TO SUBFREEZING CONDITIONS

After the blockage checks, which required perturbing the temperatures in the VCHP, the stabilization procedure was applied imposing a fixed cold source temperature ($T_{CS}$ =-80°C) and a constant heat power input ($\dot{Q}_{act}$=70 W) as the operational parameters (these can be found along with other parameters in Table 4). The VCHP was then exposed to these conditions over a time window of 211 hours, during which the temperature distribution was monitored to detect frozen blockage formation. The resulting temperature evolution over time is reported in Figure 10. The first noticeable response of the heat pipe to the subfreezing exposure is evident from the increase of $T_{adl}$ observed after about 42 hours: up to that moment, $T_{adl}$ is observed to oscillate around the steady value of -11.0°C and correlates quite well with $T_{res}$. The curve experiences a jump to -10.4°C (which is reached after about 55 hours of exposure), followed by a slow but persistent increase that continues up to 115 hours of exposure. A second jump from -10.1°C to -7.3°C is observed between 140 to 150 hours of exposure, again followed by a constant temperature increment reaching -6.7°C after 211 hours of exposure. Maintaining steady input parameters during the experiment, increments in $T_{adl}$ would not happen under normal operation conditions. In subfreezing conditions, on the other hand, $T_{adl}$ shows a transient behavior. Similar temperature increments in the evaporator and the adiabatic section of a VCHP were observed during experiments by Antoniuk et al. [35] and Edwards et al. [41], where the molecules of the working fluid diffuse into the inactive part of the condenser and they freeze in the vicinity of the ammonia freezing point location ($z_F$), causing depletion of active WF availability in the evaporator, leading to less heat dissipation, and therefore increasing the temperature in that section; they reported that diffusion is the transient mechanism behind the increase of temperature in those experiments. We assume that, in the same way, diffusion is the cause of the unsteady behavior of $T_{adl}$ in our experiment. Two more facts support this conclusion: the conditions for diffusion were met during the setup conceptualization (a big gradient between $T_{ev}$ and $T_{CS}$, and a proper VCHP temperature profile), and the time lapse before the beginning of $T_{adl}$ increments (shown in Figure 10 and Figure C 1) agrees with the time scale for diffusion of ammonia ($t_{diff} \cong O(10^5)$ s) estimated in Sec. 4.4.

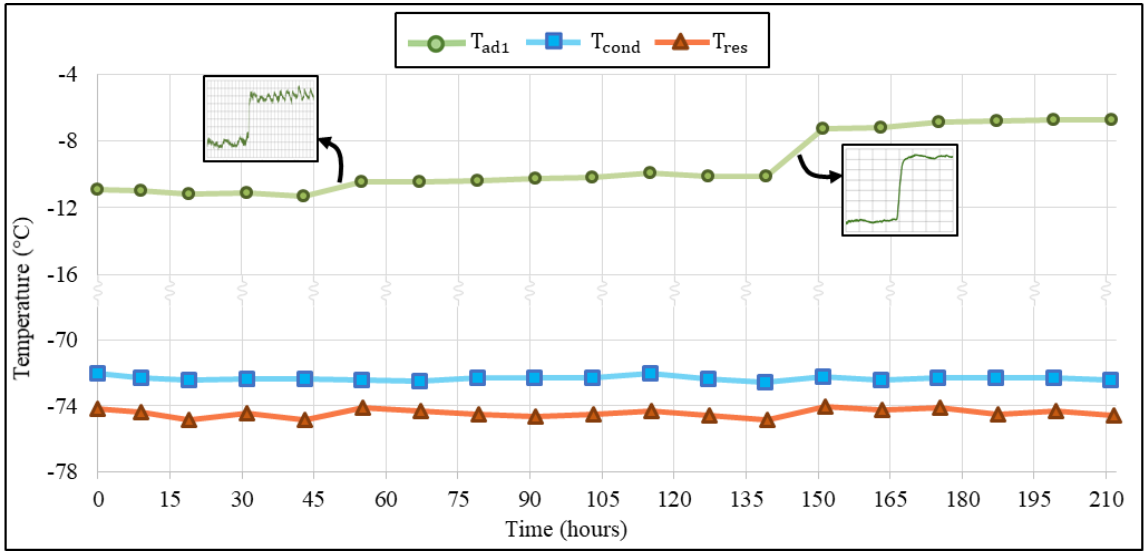

*Figure 10 VCHP long-period exposition temperatures evolution over time (for illustration purposes, data are shown every 12 hours). The dashed curve shows the average temperature in the Adiabatic 1 ($T_{ad1}$) and the close-ups (plotted at data acquisition time of 5 seconds) show in detail how this temperature increases; the solid line represents the average temperature in the condenser ($T_{cond}$) and the dotted curve shows the average temperature in the reservoir ($T_{res}$).*

### 4.6 BLOCKAGE DETECTION AND BLOCKAGE REMOVAL

As described in the last subsection, the time evolution of $T_{ad1}$ shown in Figure 10 suggests that freezing in the inactive zone and active working fluid depletion in the active zone have been acting in the VCHP since hour 48 of the testing. Therefore, the next step is to check if, at some moment after this time, a frozen layer of ammonia was formed, blocking radially the inactive core of the VCHP. The correlation between $T_{ad1}$ and $T_{res}$ was then checked, applying the procedure described in Sect. 2.3 and used in Sec. 4.3. Beginning at hour 211.6 of the testing, $T_{res}$ was cooled down smoothly from -74.3°C to -90°C over a time span of 1 hour, as shown in Figure 11. No significant change of $T_{ad1}$ was observed, indicating that the two temperatures are indeed decoupled and confirming therefore the presence of a layer of frozen ammonia covering the condenser core. Not only did we observe no decrease of $T_{ad1}$, but it was noticed to increase just before time $t = 212$ h, followed by a plateau that is maintained during the whole process. This increment can be caused by the transient condition of the system induced by the diffusion freeze-out, where the system pressure, temperature, concentration and diffusion rate are continuously evolving. To verify the origin of this behavior, we performed another shorter (90 hours) experiment, which is reported in Appendix C. The aim of this additional experiment is to show that the jump of $T_{ad1}$ during the blockage check is not triggered immediately at the start of the check, but is activated after a time lag needed by the diffused vapor molecules to reach the subfreezing temperature in the inactive part of the condenser. The reader is referred to Appendix C for further details and discussion of the obtained results.

The presence of a blockage can be also assessed by looking at the effect produced by its removal, since a frozen WF layer divides the condenser into two compartments, each having different pressure. To remove the blockage, we heated up the condenser to achieve temperatures in the inactive section above -77.78°C, i.e. above the freezing temperature of ammonia. This induces a sudden meltdown of the blockage and the resulting difference of pressures produces a wave that causes a small perturbation in the gas-vapor front, thus generating temperature fluctuations at several points along the condenser [40]. These fluctuations can be observed between time 215.7 and 215.8 hours (shown by the dotted curve in Figure 12).

After the blockage was removed, $T_{ad1}$ was found to decrease to about -10.50°C, which is very close to its stabilized value (see Table 5). Note that the temperature in the reservoir remains constant at all times.

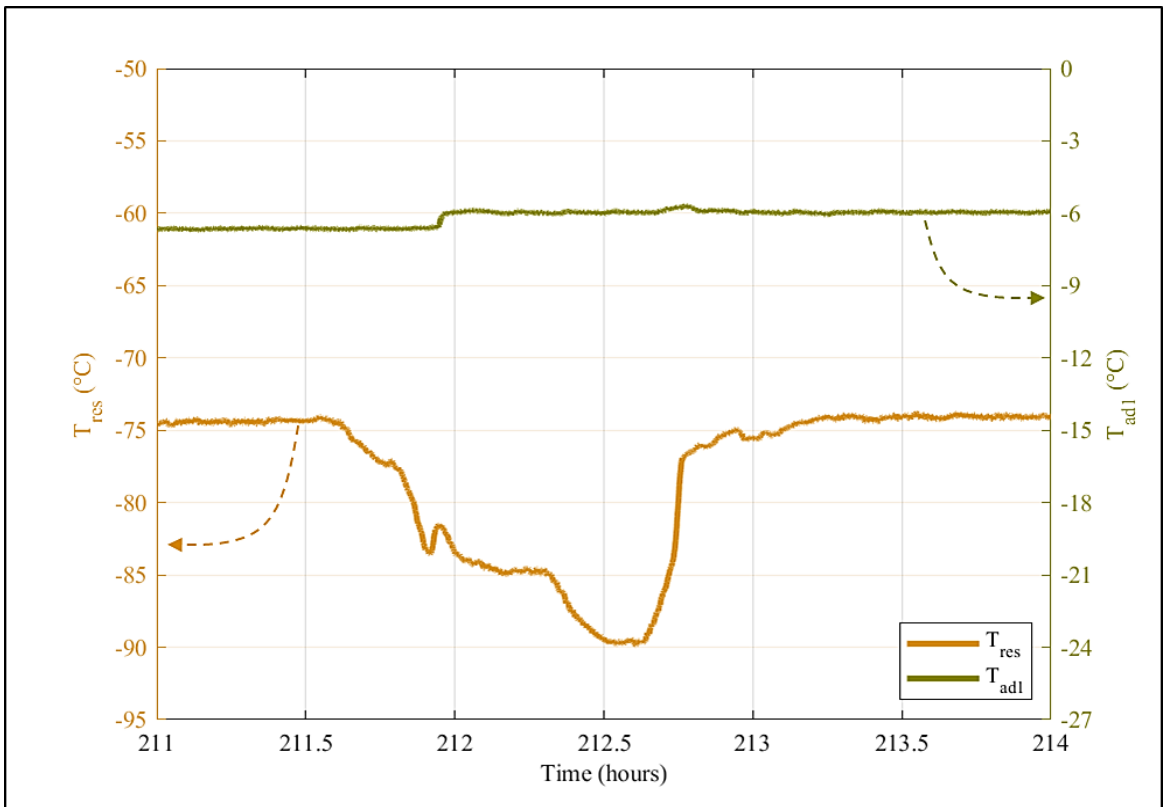

*Figure 11 Blockage check after long-period freezing exposure. Arrows have the same meaning as in Figure 8.*

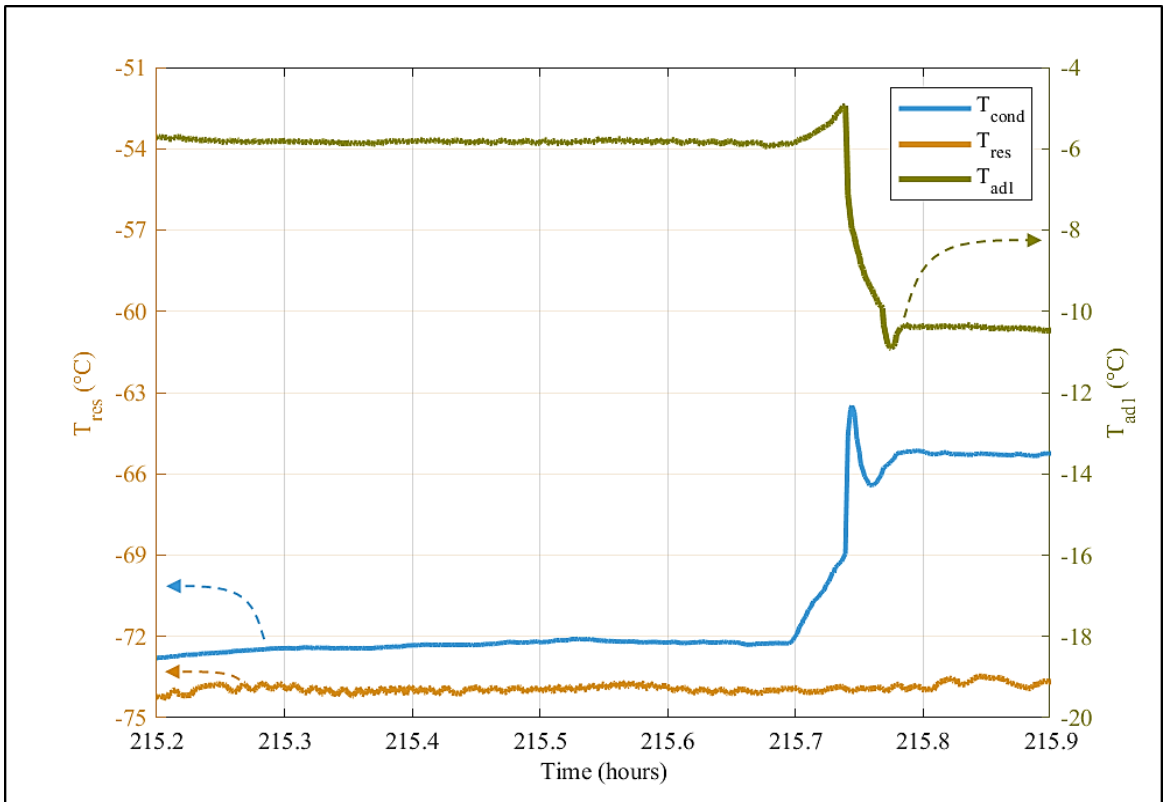

*Figure 12 Blockage rupture by heating up the condenser. Arrows have the same meaning as in Figure 8.*

To further verify the blockage removal, we performed again a blockage check by increasing $T_{res}$. The outcome of this check is evidenced in Figure 13, which clearly shows that the correlation between $T_{res}$ and $T_{ad1}$ has been re-established.

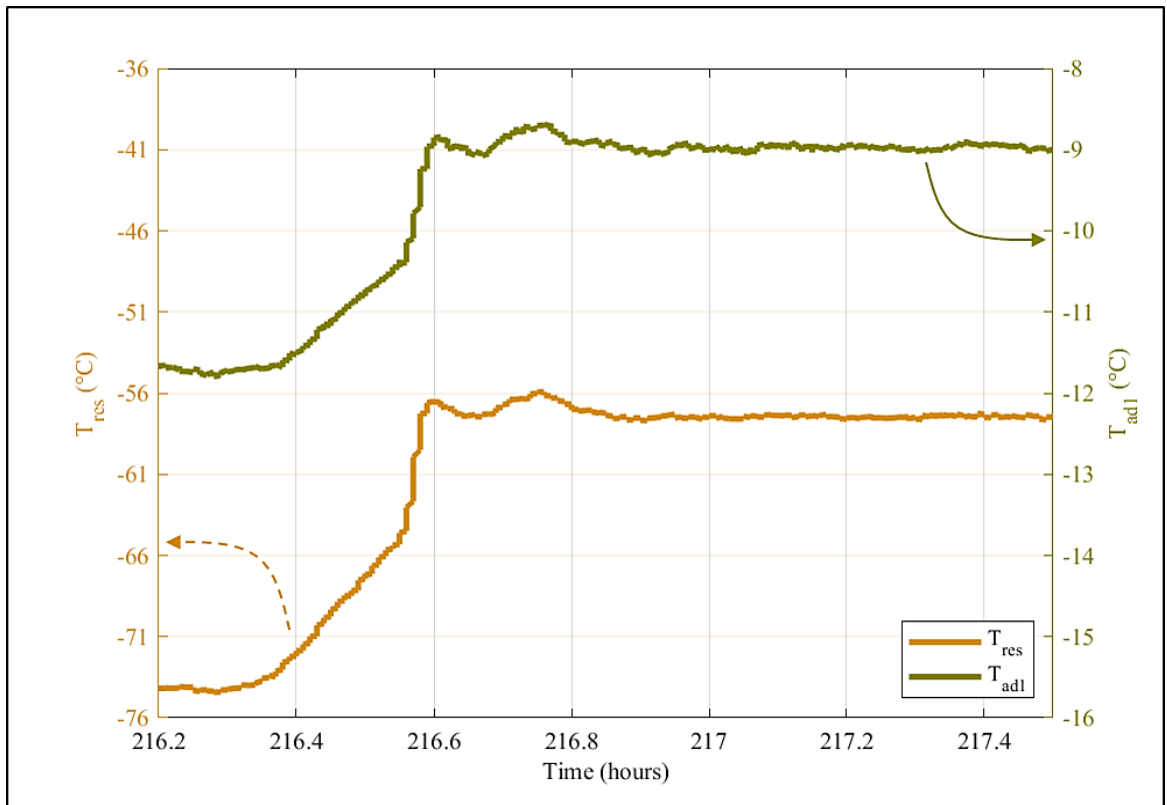

*Figure 13 Blockage check showing the re-established correlation between $T_{res}$ and $T_{ad1}$. Arrows have the same meaning as in Figure 8.*

## 4.7 PHENOMENOLOGICAL INTERPRETATION OF THE LONG-TERM EXPOSURE RESULTS

In this subsection, we elaborate on the results shown in Sec. 4.5, and in Figure 10 in particular, to attempt at an explanation of the physical processes taking place in the condenser when a frozen block is formed. With a frozen layer of ammonia blocking the heat pipe, one can envision the inactive part of the condenser divided into two sections, as shown in the diagram of Figure 14.a. Ammonia molecules can diffuse from the active part to Section I of the inactive part and, once there, move towards the non-frozen wick and be driven back towards the evaporator by the capillary forces in the wick. Section II is the subfreezing region where the ammonia molecules freeze and deposit, piling up on the frozen wick [41]. When the piling occurs, there is a loss of the active working fluid and hence less heat dissipation. This causes the increase of $T_{ad1}$ observed in Figure 10, but also produces an increase of pressure in the active part of the condenser [33], which passes from isobaric conditions at saturation pressure (equal to $P_{ad1}$), as shown schematically in Figure 14.b, to a situation in which a pressure jump occurs across the frozen plug, as shown schematically in Figure 14.c.

More specifically, without any frozen blockage, the total pressure in the heat pipe is equal to $P_{ad1}$. The partial pressure of NCG is similar to $P_{ad1}$ in the inactive part of the condenser, while dropping to much smaller values across the gas-vapor interface, while the partial pressure of ammonia increases across the gas-vapor interface, becoming similar to $P_{ad1}$ in the active part of the reservoir (see Figure 14.b) [44]. On the other hand, when a solid blockage is formed, the pressure distributions of ammonia and of NCG develop as shown in Figure 14.c. In Section II, the partial pressures of both ammonia and NCG are not altered significantly, compared with the previous case, but undergo an increase across the frozen block. In Section I, they change continuously based on a new total pressure in the active part, namely $P_{ad1}$. This value is greater than $P_{ad1}$ in the absence of a blockage, due to the growth of $T_{ad1}$ during the blockage formation shown in Figure 10, generating thus a total pressure difference in the condenser observed as a step-like jump across the frozen blockage. This gradient of pressure enhances the diffusion of vapor ammonia molecules from the active to the inactive side of the condenser, through the gas-vapor front. Because of this, the diffusive flux of ammonia across Section I is increased and more ammonia molecules can reach the freezing point, thus providing a self-triggering mechanism for the diffusive growth of the frozen blockage. We point out that as long as this mechanism is active, the concentration of ammonia in the active part decreases, leading to dry-out and causing further malfunctioning [40].

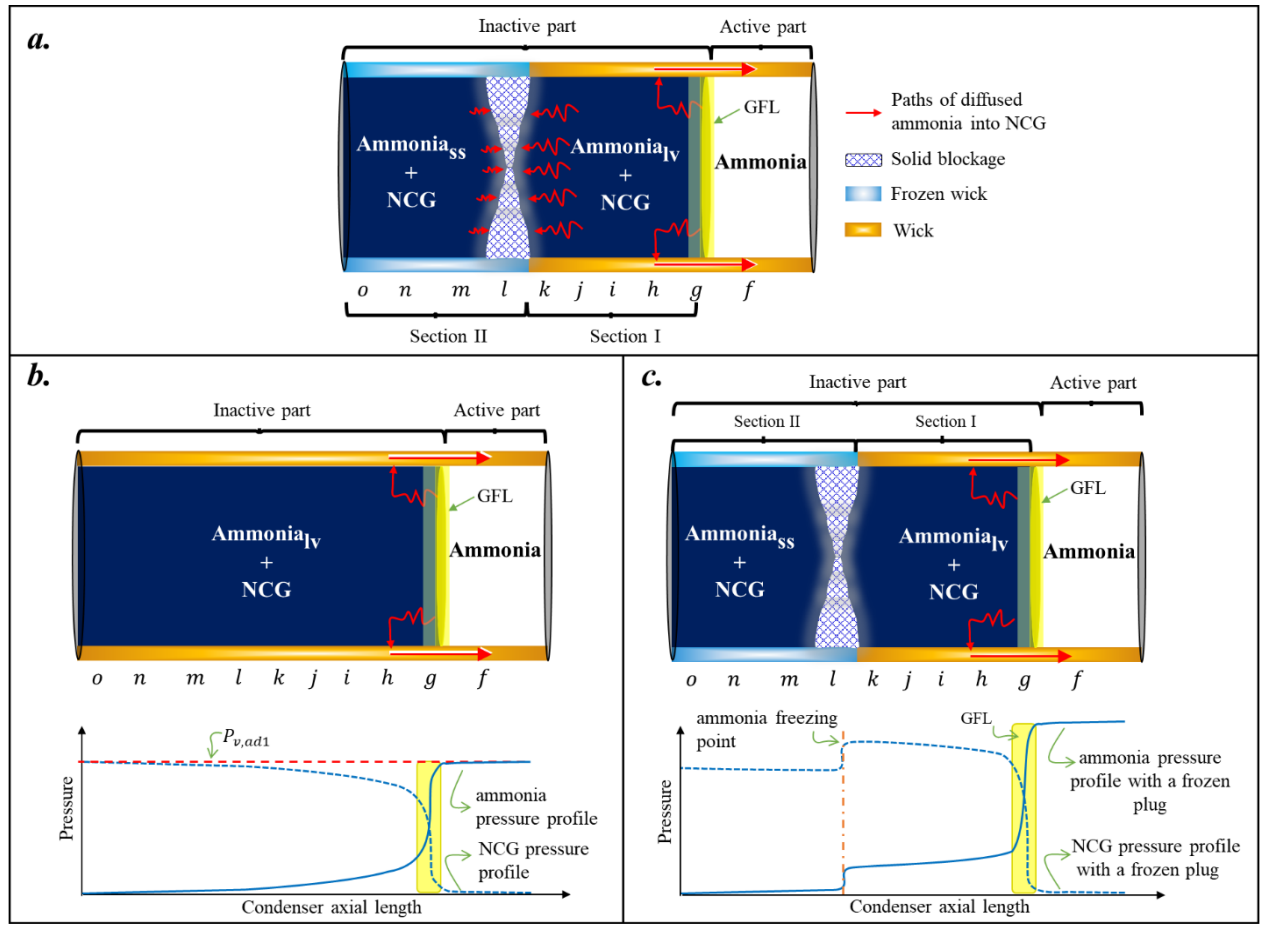

*Figure 14  **a**. Diagram of diffused vapor ammonia paths within the inactive part of the condenser. **b**. Schematic representation of the pressure distribution along the condenser in the absence of a blockage. The total pressure is uniform throughout the VCHP. **c**. Diagram of ammonia and NCG pressure profiles in the condenser after a frozen blockage is formed. In panels **a** and **c**, sublimated and solid ammonia coexist in the left region at saturation pressure conditions [46]. Hence, the subscript "ss" is used to identify sublimated-solid ammonia (below the freezing point), while the subscript "lv" is used to identify liquid-vapor ammonia (above the freezing point).*

## 5 CONCLUSIONS

In this work, a computational routine based on the modified Flat Front Approach was implemented and verified experimentally with 8 cases under different operating conditions. This approach allowed us to compute the location of the gas-vapor front (GLF) and to predict the VCHP temperature profile. Knowing the temperature distribution along the VCHP for different conditions helped us conceive the design of the experimental setup and the procedure for achieving the formation of a frozen ammonia layer blocking radially the inactive part of an ammonia/stainless-steel VCHP and for its successful detection.

Using a test bench, the VCHP was set up in a way that the temperature in the reservoir ($T_{res}$) and in the condenser ($T_{cond}$) could be controlled using heaters and cold plates (thermal exchangers using liquid nitrogen). The whole setup was isolated using polystyrene beads. After performing health and calibration checks of the setup, subfreezing experiments on the VCHP were performed following the test sequence shown in Table 2. A detection technique based on the measurement of the correlated $T_{res}/T_{ad1}$ profiles was developed to estimate the formation (or absence) of the frozen blockage. The correlation between $T_{res}$ and $T_{ad1}$ turned out to be very strong for the operation conditions tested in the calibration step, and was clearly broken after a long-enough period (of about 211 hours) of VCHP exposure to subfreezing temperatures.

During the freezing experiment, the temperatures $T_{res}$ and $T_{CS}$, as well as the heat power input were kept constant. However, an evident increase of temperature $T_{ad1}$ in time could be observed. This change of temperature was attributed to the depletion of the active working fluid by diffusion and freezing in the inactive part of the condenser. Devices like external heaters attached to the VCHP condenser can melt an ammonia frozen block formed inside the heat pipe, similarly to the melt down test we performed (where $T_{ad1}$ returned to its initial value and the correlation between $T_{res}$ and $T_{ad1}$ was re-established).

The results of our study show that in the environment where the VCHP operates, a frozen blockage may form, proving thus that current-adopted practices such as mounting heating devices in the VCHP condenser to avoid frozen blockage formation are not only preventive, but justified from a scientific perspective and essential. They also prove that diffusion is a very important phenomenon that can influence the normal operation of VCHPs and should be carefully considered, in order to ensure a proper design of ammonia/stainless-steel VCHPs exposed to subfreezing conditions.

**Declaration of competing interest**
The authors declare that they have no known competing financial interests or personal relationships that could have appeared to influence the work reported in this paper.

**Data availability**
Data will be made available on request.

**Acknowledgements**
F. K. Miranda gratefully acknowledges EHP for the support provided during the industrial secondment. This work has received funding from the European Union's Horizon 2020 research and innovation programme under Marie Skłodowska-Curie grant agreement no. 813948 (COMETE).

# APPENDIX A

## Validation experiments

We performed a total of eight experimental case studies (see Figure A 1) to validate the computational routine stemming from equations (1) to (6). Each validation case was carried out using different input values for $T_{res}$, $T_{cs}$, and $\dot{Q}_{in}$, in order to vary the gas-vapor front location (GFL), which is determined by the mathematical model and can be also visually located at the end of the temperature plateau in the VCHP temperature profile.

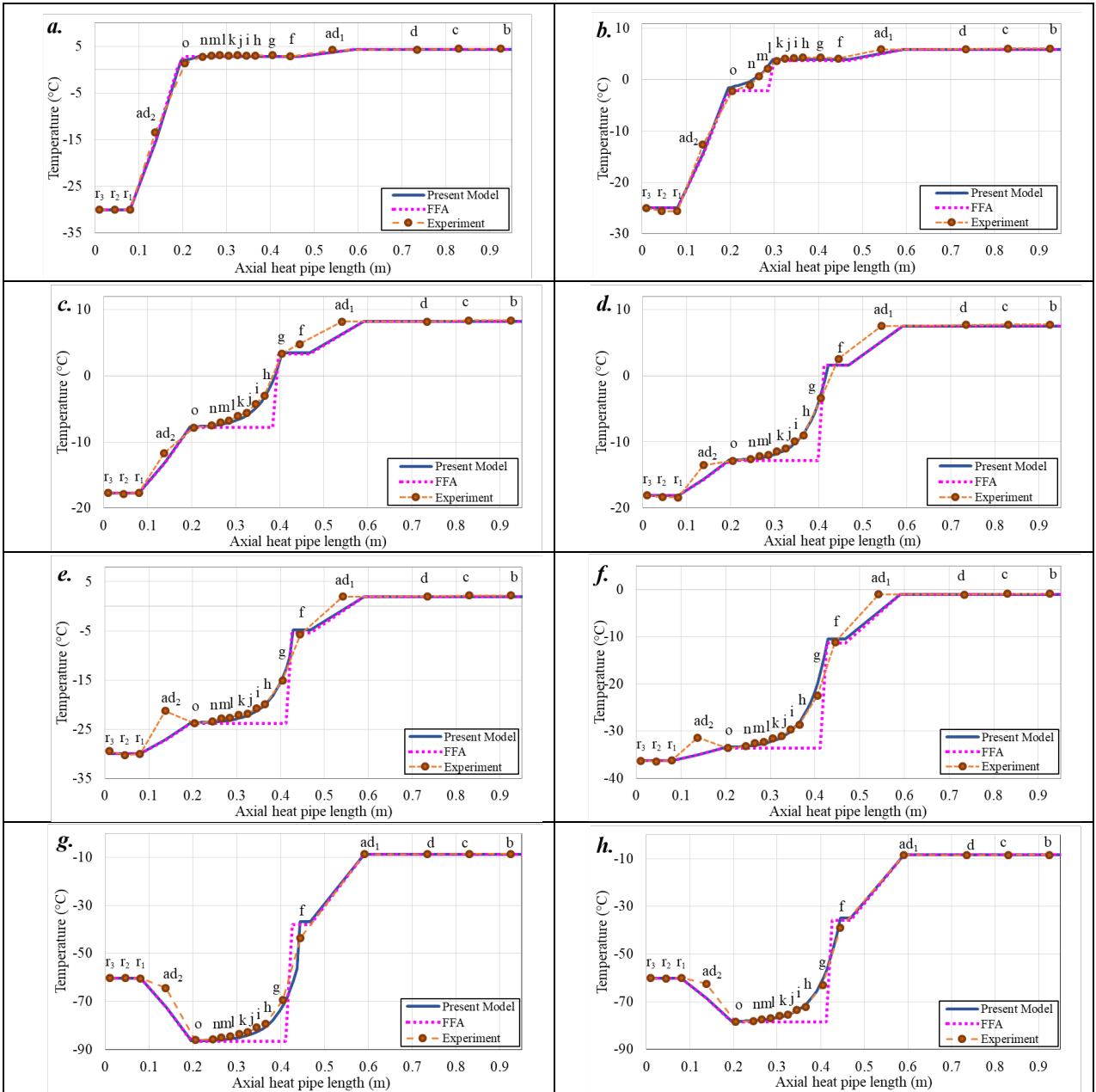

Figure A 1 **a. Fully open VHCP** using $T_{res} = -29.80°C$, $T_{cs} = 0.30°C$ and $T_{ad1} = 4.28°C$ with input heat power of $\dot{Q}_{in} = 25\ W$. **b. Half open VCHP** using $T_{res} = -25.00°C$, $T_{cs} = -2.17°C$ and $T_{ad1} = 5.94°C$ with an input power of $\dot{Q}_{in} = 25\ W$. **c. One-third open VCHP** using $T_{res} = -17.70°C$, $T_{cs} = -7.82°C$ and $T_{ad1} = 8.20°C$ with an input power of $\dot{Q}_{in} = 25\ W$. **d. One-third open VCHP** using $T_{res} = -18.10°C$, $T_{cs} = -12.86°C$ and $T_{ad1} = 7.50°C$ with an input power of $\dot{Q}_{in} = 25\ W$. **e. One-third open VCHP** using $T_{res} = -30.00°C$, $T_{cs} = -32.77°C$ and $T_{ad1} = 1.93°C$ with an input power of $\dot{Q}_{in} = 25\ W$. **f. One-third open VCHP** using $T_{res} = -36.30°C$, $T_{cs} = -33.60°C$ and $T_{ad1} = -1.02°C$ with an input power of $\dot{Q}_{in} = 35\ W$. **g. One-third open VCHP** using $T_{res} = -60.30°C$, $T_{cs} = -86.60°C$ and $T_{ad1} = -8.70°C$ with an input power of $\dot{Q}_{in} = 75\ W$. **h. One-third open VCHP** using $T_{res} = -60.10°C$, $T_{cs} = -78.45°C$ and $T_{ad1} = -8.40°C$ with an input power of $\dot{Q}_{in} = 70\ W$.

Panel "*a*" shows a fully open condenser with the GFL positioned around point "o" and a reservoir temperature colder than the cold source. Panel "*b*" depicts a half open condenser with the GFL positioned around point "*l*" with the reservoir colder than the cold source. Panels "*c*", "*d*", "*e*" and "*f*" refer to the case of a condenser that is one-third open with the GFL positioned around points "*g*" and "*f*" and the reservoir temperature approaching the cold source temperature, but still colder. Panels "*g*" and "*h*" refer to the case of a condenser that is one-third open with the GFL positioned around point "*f*", with the reservoir hotter than the cold source, with the temperatures below the freezing point of the working fluid (ammonia).

## APPENDIX B

### Fin equation derivation

Let us consider a long fin with its axis in $x$ direction, subject to convective cooling along its outer surface, and assume that conduction phenomenon occurs in the axial direction only. Taking into account a differential element as shown in the right side of Figure B 1, we can assume that the incoming heat $q_x$ may change after a distance $dx$ because of the heat loss $dq_{ht}$ that occurs through the lateral surface area $dA_S$. Using Taylor expansion to express the outgoing flux as:

$$q_{x+dx} = q_x + \frac{dq_x}{dx} dx, \tag{B1}$$

using the energy balance:

$$q_{x+dx} = q_x - dq_{ht}, \tag{B2}$$

and considering Fourier's law of heat conduction with $k$ the thermal conductivity and $A_C$ the cross-sectional area, one gets:

$$q_x = -kA_C \frac{dT}{dx}, \tag{B3}$$

namely Newton's law of cooling with $HTC$ as the heat dissipation coefficient and $A_S$ as the external surface area of the fin:

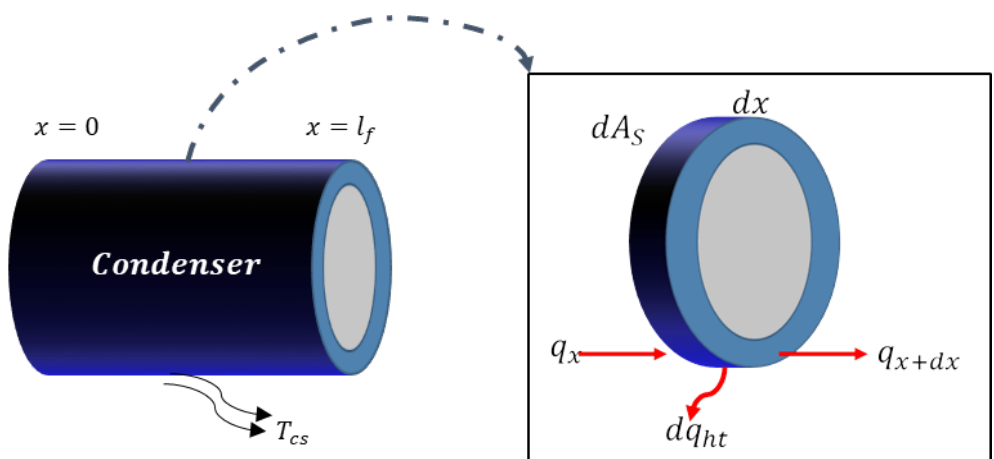

*Figure B 1 Fin differential element analysis.*

$$dq_{ht} = HTC \, dA_S (T - T_{cs}) . \tag{B4}$$

Incorporating equations (B2), (B3) and (B4) into equation (B1) and expressing the external surface area as $dA_S = \mathcal{P} dx$ with $\mathcal{P}$ the perimeter of the differential element, we have:

$$-kA_C \frac{dT}{dx} = -kA_C \frac{dT}{dx} - kA_C \frac{d^2T}{dx^2} dx + HTC \, \mathcal{P} dx (T - T_{cs}) . \tag{B5}$$

This yields the general differential equation for the fin:

$$kA_C \frac{d^2T}{dx^2} dx = HTC \, \mathcal{P} \, dx (T - T_{cs}) . \tag{B6}$$

To keep the solution simple, we introduce $\theta = T - T_{cs}$ with its derivative $\theta' = T'_{cs}$ and $m = \text{sqrt}\left(\frac{HTC \, \mathcal{P}}{kA_C}\right)$ into equation (B6) to get:

$$\theta'' - m^2 \theta = 0 , \tag{B7}$$

which has a well-known general solution of the form:

$$\theta = C_1 e^{mx} . \tag{B8}$$

The value of $C_1$ can be found using the known values of $x$ and $T$ at the freezing point, i.e. $x = l_f$ and $T = T_F$:

$$C_1 = (T_F - T_{cs}) e^{-ml_f} , \tag{B9}$$

incorporating it into equation (B8):

$$\theta = (T_F - T_{cs}) e^{m(x - l_f)} , \tag{B10}$$

where $l_f$ is the axial position of the freezing point in the condenser and $T_f$ is the freezing temperature of the working fluid. Taking the derivative of equation B10 with respect to the axial variable $x$ and evaluating it at the freezing axial location, one gets:

$$\frac{dT}{dx} = m(T_F - T_{cs}) . \tag{B11}$$

**APPENDIX C**

### Adiabatic temperature increment over time

To examine the time evolution of the temperature in the Adiabatic 1 ($T_{ad1}$) and show that it does not start increasing immediately after the start of the experiment, we carried out a test case in which the temperature profile shown in Table C 1 was used and a constant heat power input of 70 W was imposed. Like the experiments of long exposure to subfreezing temperatures, the temperatures $T_{res}$ and $T_{cs}$ were also kept constant during the whole experiment. Similar to what has already been shown in Figure 10, it is observed that $T_{ad1}$ exhibits a transient evolution (shown in Figure C 1), and it begins increasing only after 27 hours from the start of the experiment. This increase is activated after a time that corresponds to the time needed by the vapor molecules to reach by diffusion the subfreezing temperatures in the inactive part of the condenser.

Table C 1 Initial temperature profile measurements, along with the standard deviation (SD), the standard error of mean (SEM), the upper and lower limit of the confidence interval 95%, (CI 95% u and CI 95% l).

| Point | Mean (°C) | SD (°C) | SEM (°C) | CI 95% u (°C) | CI 95% l (°C) |
|---|---|---|---|---|---|
| $T_{res}$ | -74.30 | 0.25 | 0.0559 | -74.1904 | -74.4096 |
| $T_o$ | -80,01 | 0.18 | 0.0402 | -79.9311 | -80.0889 |
| $T_l$ | -78,35 | 0.18 | 0.0402 | -78.2711 | -78.4289 |
| $T_k$ | -77, 31 | 0.17 | 0.0380 | -77.2354 | -77.3845 |
| $T_j$ | -76.63 | 0.16 | 0.0357 | -76.5598 | -76.7001 |
| $T_i$ | -74.83 | 0.17 | 0.0380 | -74.7554 | -74.9045 |
| $T_h$ | -73.37 | 0.16 | 0.0357 | -73.2998 | -73.4401 |
| $T_g$ | -64.16 | 0.18 | 0.0402 | -64.0811 | -64.2389 |
| $T_f$ | -39.89 | 0.16 | 0.0357 | -39.8198 | -39.9601 |
| $T_{ad1}$ | -8.84 | 0.10 | 0.0223 | -8.7961 | -8.8838 |

Focusing on the behavior of $T_{ad1}$, represented by the dashed curve in Figure C 1, it is easy to observe that a constant value of about -8.67°C is kept for the first 25 hours of the experiment. Subsequently, $T_{ad1}$ experiences a first jump up to -8.45°C, followed by a small increment up to -8.10°C until the time $t = 45$ from the start of the experiment. Finally, a large increment up to -7.20 °C is observed after 70 hours from the start of this experiment.

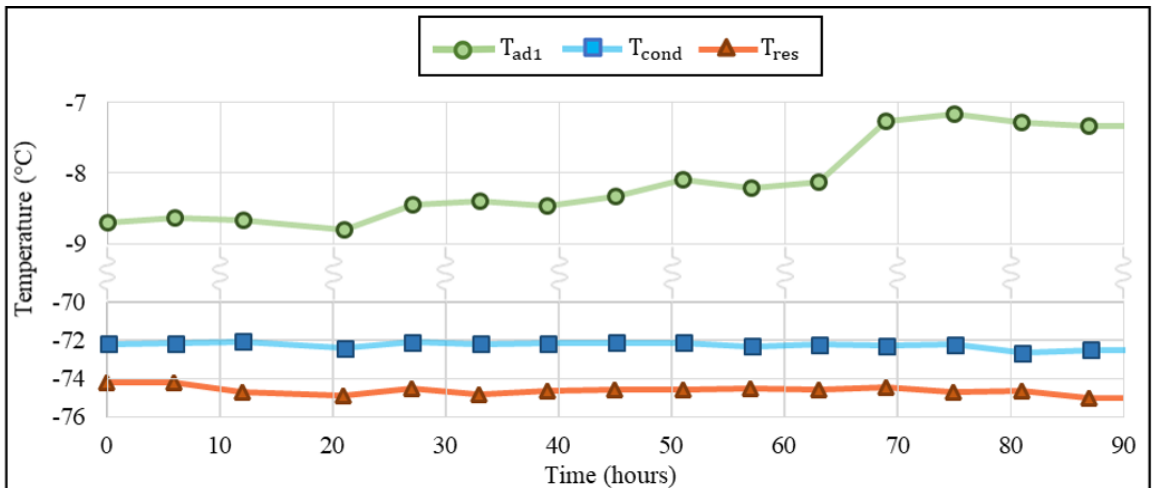

Figure C 1 Diffusion-freezing activation time.